\documentclass{jfm}

\usepackage{eqnarray,amsmath}

\usepackage{subcaption}
\usepackage{hyperref}
\newcommand{\comments}[1]{}

\usepackage{graphicx}
\usepackage{amssymb}

\usepackage{natbib}
\usepackage{rotating}

\newcommand{\bddn}[1]{\frac{\partial #1}{\partial n}}
\newcommand{\bma}[1]{\mbox{$\bf #1$}}
\newcommand{\vnabla}{\bma{\nabla}}

\title[A unified breaking onset criterion for surface gravity water waves]{A unified breaking onset criterion for surface gravity water waves in arbitrary depth}

\author[Derakhti et al.]{Morteza Derakhti$^1$ \thanks{Email address for correspondence: derakhti@uw.edu}, James T. Kirby$^2$, Michael L. Banner$^3$, Stephan T. Grilli$^4$ \and \; Jim Thomson$^1$}

\affiliation{$^1$ Applied Physics Lab, University of Washington, Seattle, WA, USA \\
$^2$  Center for Applied Coastal Research, University of Delaware, Newark, DE 19716, USA \\
$^3$  School of Mathematics and Statistics, University of New South Wales, Sydney, Australia \\
$^4$  Department of Ocean Engineering, University of Rhode Island, Narragansett, RI, USA }

\begin{document}

\maketitle

\begin{abstract}

We investigate the validity and robustness of the \citet{Barthelemy-etal:2018} breaking wave onset prediction framework for surface gravity water waves in arbitrary water depth, including shallow water breaking over varying bathymetry. We show that the \citet{Barthelemy-etal:2018} breaking onset criterion, which they validated for deep and intermediate water depths, also segregates breaking crests from non-breaking crests in shallow water, with subsequent breaking always following the exceedance of their proposed generic breaking threshold. We consider a number of representative wave types, including regular, irregular, solitary, and focused waves, shoaling over idealized bed topographies including an idealized bar geometry and a mildly- to steeply-sloping planar beach. Our results show that the new breaking onset criterion is capable of detecting single and multiple breaking events in time and space in arbitrary water depth. Further, we show that the new generic criterion provides improved skill for signaling imminent breaking onset, relative to the available kinematic or geometric breaking onset criteria in the literature. In particular, the new criterion is suitable for use in wave-resolving models that cannot intrinsically detect the onset of wave breaking.

\end{abstract}

\section{Introduction} \label{S:1}

Finding a robust and universal diagnostic parameter that determines the onset of breaking and its strength is of substantial importance in the prediction of atmosphere-ocean exchanges, nearshore circulation and mixing, design of offshore and nearshore infrastructures, etc, but as yet the problem is not completely resolved. 

Considerable effort has been made to find a robust and universal methodology to predict the onset of breaking gravity water waves in deep and intermediate depth water \citep{song-banner-jpo02,Wu-Nepf:2002,Banner-Peirson:2007,Babanin-etal:2007,tian-etal-pf08,Toffoli-etal:2010,Shemer-Liberzon:2014,Fedele-etal:2016,Saket-etal:2017,Saket-etal:2018,Barthelemy-etal:2018,Khait-Shemer:2018,Craciunescu-Christou:2019:identifying,pizzo-melville-jfm19}. 
This and other aspects of wave breaking have been covered in several excellent reviews of the topic \citep{Banner-Peregrine:1993,Melville:1996,Perlin-etal:2013}. Recently, \citet{Perlin-etal:2013} have reviewed the latest progress on prediction of geometry, breaking onset, and energy dissipation of steepness-limited breaking waves. The predictive parameters involved can be categorized as (i) geometric, (ii) kinematic, and (iii) dynamic criteria. As summarized in \citet[][\S3]{Perlin-etal:2013}, none of the available criteria can distinguish between breaking and non-breaking crests in a universal sense. 

The situation becomes even more complex in shallow water, where waves evolve in response to interaction with seabeds of arbitrary, complex geometry.  The importance of water depth $d$ as a limiting factor for shallow water breaking leads to the identification of a convenient dimensionless parameter $\gamma = H/d$ \citep{mccowan-philmag94}, where $H$ is the local wave height. Further, analysis of breaking criteria for the simplest case of waves shoaling over a planar slope introduces the slope itself as a parameter.  The effect of bottom slope $m$ in combination with a measure of wave steepness has been studied by \citet{iribarren-nogales-1949}, who defined a single combination $\xi_0 = m/\sqrt{H_0/L_0}$ based on offshore wave height $H_0$ and wave length $L_0$, and \citet{battjes-icce74}, who defined a similar surf similarity parameter $\xi_b = m/\sqrt{H_b/L_b}$, with the index $b$ denoting values taken at the breaking onset.  The surf similarity parameter has been found to be useful in discriminating between breaker types as well as in refining the prediction of breaking onset based on $\gamma$.  The range of results in the literature is reviewed by \citet{Robertson-etal:2013}, who list six types of dependency of $\gamma_b$ on additional parameters such as $m$ and $\xi_0$, and provide a table of thirty six examples of published formulae for the estimation of $\gamma_b$. \citeauthor{Robertson-etal:2013} concluded that a single, easily implementable relationship covering all breaking phenomena is still elusive.

Based on numerical simulation of 2D and 3D focused wave packets in deep and intermediate depths, \citet{Barthelemy-etal:2018} showed that the highest non-breaking waves in deep and intermediate water were clearly separated from marginally breaking waves by the value of the normalized energy flux localized near the crest tip region, and that initial breaking instability occurs within a very compact region centered on the wave crest.  On the free surface, the expression for normalized energy flux (denoted by symbol $B$) reduces to the ratio of liquid velocity at the crest $U$ in the direction of propagation to the translational velocity $C$ of the crest for the tallest wave in the evolving group. 
\citet{Barthelemy-etal:2018} found that a value of $B\approx0.85$ provides a robust threshold as a precursor to breaking for 2D wave packets propagating in deep or intermediate uniform water depths. Further, a targeted study of representative cases of the most severe laterally-focused 3D wave packets in deep and intermediate depth water shows that the threshold remains robust. Subsequently, using a different modeling framework, \citet{Derakhti-etal:2018} found consistent results for representative cases of modulated wave trains and focused packets in deep and intermediate depth water. These numerical findings for 2D and 3D cases were closely supported by the laboratory experiments of \citet{Saket-etal:2017,Saket-etal:2018}. 

It remains to be seen whether the criterion proposed by \citeauthor{Barthelemy-etal:2018} has applicability to waves in shallow water with relatively rapidly varying depth.  Our goal here is to describe a robust and local criterion that predicts the onset of breaking for gravity water waves in shallow water, extending the intermediate and deep-water results of \citet{Barthelemy-etal:2018} and \citet{Derakhti-etal:2018} to cover all water depths. The utility of a precursor such as $B=0.85$, rather than the classic $B=1$ during breaking, is the application to models that cannot directly resolve breaking and fail before waves reach $B=1$. 
We use a large-eddy-simulation (LES)/volume-of-fluid (VOF) model \citep{Derakhti-Kirby:2014, Derakhti-Kirby:2016} and a 2D fully nonlinear potential flow solver using a boundary element method (FNPF-BEM) \citep{grilli1989efficient, grilli1996numerical} to simulate nonlinear wave evolution, focusing on breaking onset behavior.  
Simulations are conducted for a variety of scenarios including regular, irregular, solitary, and focused waves shoaling over idealized bed topographies, including an idealized bar geometry and mildly- to steeply-sloping planar beaches.
Additionally, we examine the applicability of the criterion for surging/collapsing breaking cases in shallow water, for which an instability leading to breaking may develop close to the toe of the wave front. 

\section{Computational approaches}\label{S:2}

In this section, we provide a brief overview of the two modeling approaches used: the polydisperse two-fluid LES/VOF model of \citet{Derakhti-Kirby:2014} based on the model TRUCHAS \citep{Francois-etal:2006:TRUCHAS}, and the fully nonlinear potential flow - boundary element model (FNPF-BEM) model of \citet{grilli1989efficient} and \citet{grilli1996numerical}.  The cases considered here are essentially 2D in the $(x,z)$ plane, allowing us to employ a purely 2D version of FNPF-BEM.  
The FNPF-BEM model is not valid beyond the first onset of breaking, and is thus only used below to consider the transient solitary wave cases.

Validation of the models for the present application is discussed in Appendix A.

\subsection{The LES/VOF model}\label{sec2.1}

The LES/VOF computations are performed using the Navier-Stokes solver TRUCHAS \citep{Francois-etal:2006:TRUCHAS} with extensions of a polydisperse bubble phase and various turbulence models \citep{Carrica_etal:1999, ma-etal-jgrc11, Derakhti-Kirby:2014}.  
Details of the current mathematical formulations and numerical methods may be found in \citet[][\S2]{Derakhti-Kirby:2014}. 

The filtered governing equations for conservation of mass and momentum of the liquid phase are given by:
\begin{eqnarray}
\frac{\partial \alpha \rho}{\partial t} + \frac{\partial \alpha \rho \tilde{u}_{j}}{\partial x_j} = 0, \label{eq1} \\
\frac{\partial \alpha \rho \tilde{u}_{i}}{\partial t} + \frac{\partial \alpha \rho \tilde{u}_{i} \tilde{u}_{j}}{\partial x_j} = \frac{\partial \Pi_{ij}}{\partial x_j} + \alpha \rho {g} \delta_{3i} + {\bf M}^{gl}, \label{eq2}
\end{eqnarray}
where $(i,j) = 1, 2, 3$; $\rho$ is a constant liquid density; $\alpha$ and $\tilde{u}_{i}$ are the volume fraction and the filtered velocity in the $i$ direction of the liquid phase, respectively; $\delta_{ij}$ is the Kronecher delta function; $g$ is the gravitational acceleration; and $\Pi_{ij} = \alpha(-\tilde{p}\delta_{ij}+\tilde\sigma_{ij}  -\tau_{ij})$ with $\tilde{p}$ the filtered pressure, which is identical in each phase due to the neglect of interfacial surface tension, $\tilde\sigma_{ij}$ viscous stress and $\tau_{ij}$ the subgrid-scale (SGS) stress estimated using an eddy viscosity assumption and the Dynamic Smagorinsky model, which includes water/bubble interaction effects \citep[for more details see][\S 2.4]{Derakhti-Kirby:2014}. Finally, ${\bf M}^{gl}$ are the momentum transfers between liquid and gas phases, including the filtered virtual mass, lift, and drag forces \citep[][\S 2.2]{Derakhti-Kirby:2014}.

Using the same filtering process as in the liquid phase, the equations for the bubble number density and continuity of momentum for each bubble size class with a diameter $d_k^b$, $k=1,\cdots,N_G$, are then given by \citep[][\S 2]{Derakhti-Kirby:2014}:
\begin{eqnarray}
\frac{\partial N_{k}^b}{\partial t} + \frac{\partial \tilde{u}_{k,j}^b N_{k}^b}{\partial x_j} = {R}_{k}^b, \label{eq3} \\
0 = -\frac{\partial \alpha_k^b \tilde{p}}{\partial x_j}\delta_{ij} + \alpha_{k}^b \rho^b {g_i} + {\bf M}_{k}^{lg}, \label{eq4}
\end{eqnarray}
where $\alpha^b_k = m^b_k N_{k}^b/\rho^b$, $m^b_k$, $N_{k}^b$ and $\tilde{u}_{k,j}^b$ are the volume fraction, mass, number density and filtered velocity in the $j$ direction of the $k$th bubble size class; $\rho^b$ is the bubble density; and $R^b_k$ includes the source due to air entrainment in the interfacial cells \citep[][\S 2.3]{Derakhti-Kirby:2014}, intergroup mass transfer, and SGS diffusion terms. Finally, ${\bf M}_{k}^{lg}$ represents the total momentum transfer between liquid and the $k$th bubble size class, and satisfies ${\bf M}^{gl} + \sum_{k=1}^{N_G} {\bf M}_{k}^{lg} = 0$.  In (\ref{eq4}), we neglect the inertia and shear stress terms in the gas phase following \citet{Carrica_etal:1999} and \citet{Derakhti-Kirby:2014}.

\subsection{The FNPF-BEM model} \label{sec2.2}

Equations for the two-dimensional (2D) FNPF-BEM model are briefly presented here.  The velocity potential $\phi(\bma{x},t)$ is used to describe inviscid, irrotational flow in the vertical plane $(x,z)$, with the velocity defined by $\bma{u} = \vnabla\phi = (u,w)$.
$\phi$ is governed by Laplace's equation in the liquid domain $\Omega(t)$ with boundary $\Gamma(t)$,
\begin{equation}
\nabla^2 \phi = 0; \hspace{0.25in}  (x,z) \in \Omega(t)  \label{eq5}
\end{equation}
Using the 2-D free space Green's function, $G(\bma{x},\bma{x}_l)=-(1/2\pi)\log{ \mid \bma{x-x}_l \mid}$, and Green's second identity, (\ref{eq5}) is transformed into the boundary integral equation
\begin{equation}
\alpha(\bma{x}_l)\phi(\bma{x}_l) = \int_{\Gamma(\bma{x})} 
 [{\bddn{\phi}}(\bma{x}) G(\bma{x},\bma{x}_l) - \phi(\bma{x}) 
 \bddn{G(\bma{x},\bma{x}_l)}]\, d\Gamma(\bma{x})          \label{eq6} 
\end{equation}
where $\bma{x}=(x,z)$ and $\bma{x}_l=(x_l,z_l)$ are position vectors for points on the boundary, $\bma{n}$ is the unit outward normal vector, and $\alpha(\bma{x}_l)$ is a geometric coefficient. Details of the surface and bottom boundary conditions and numerical methods may be found in \citet{grilli1989efficient} and \citet{grilli1996numerical}. The model provides instantaneous surface elevation and liquid velocity at the surface. 

\section{Model configuration and test cases} \label{sec3}

\subsection{Definition of breaking} \label{sec3.1}

We consider an individual crest to be a breaking crest if 1) a multi-valued free surface forms at the crest, or 2) energy flux loss exceeds background viscous dissipation. In the BEM framework, there is no dissipation mechanism in the model, and the model becomes unstable fairly rapidly after a vertical tangent becomes apparent at the crest. Thus a crest in simulations using the FNPF-BEM model is denoted a non-breaking crest if its maximum surface elevation reaches a maximum and then decreases. 

\subsection{Test cases} \label{sec3.2}

\begin{figure}
\centering
\includegraphics[width=1\textwidth]{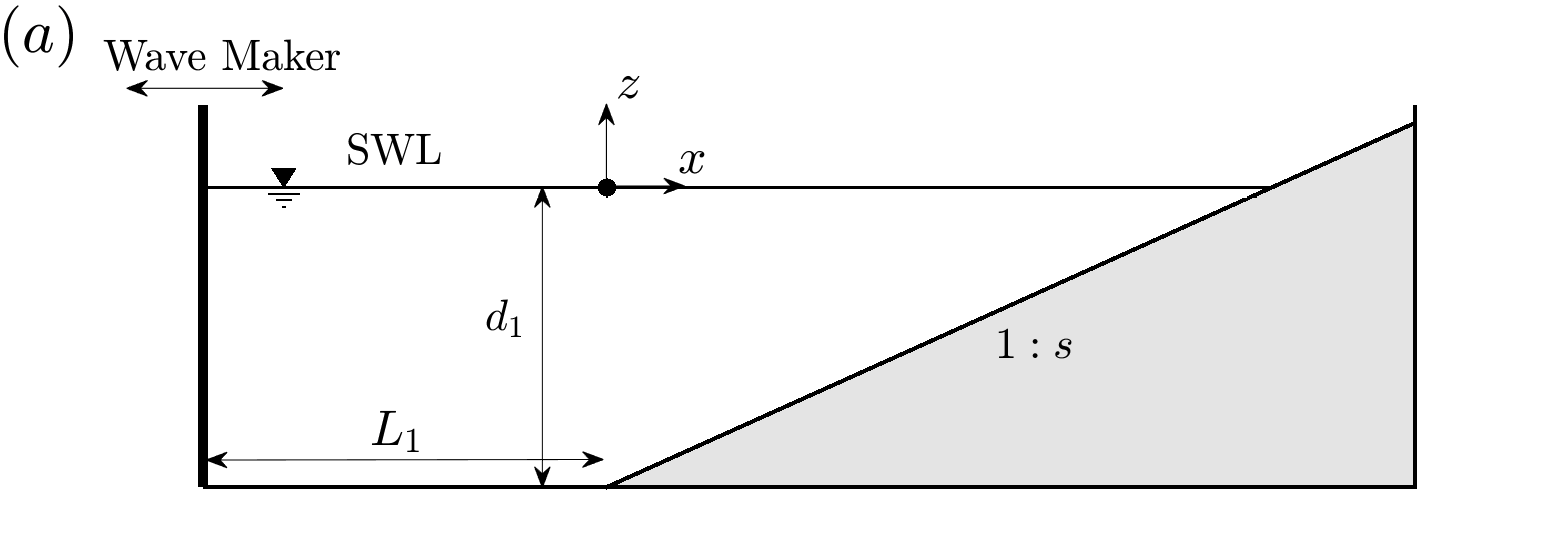}
\includegraphics[width=1\textwidth]{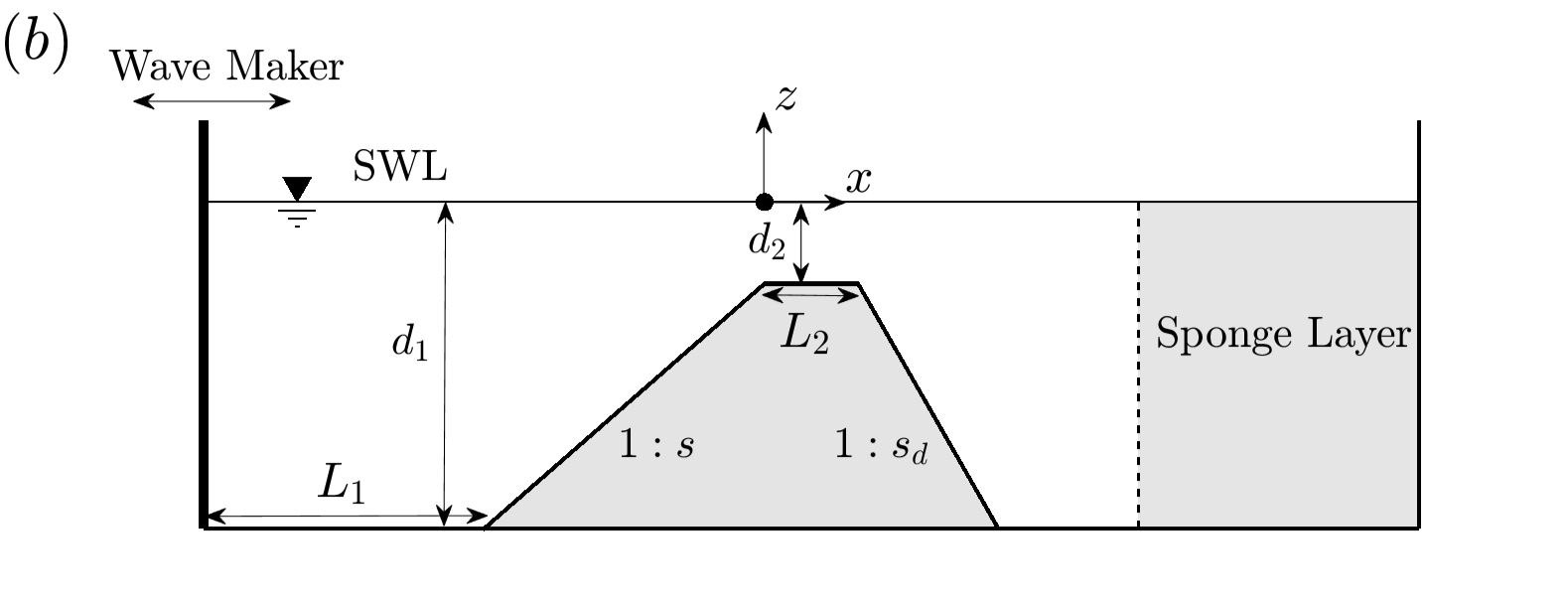}
\includegraphics[width=1\textwidth]{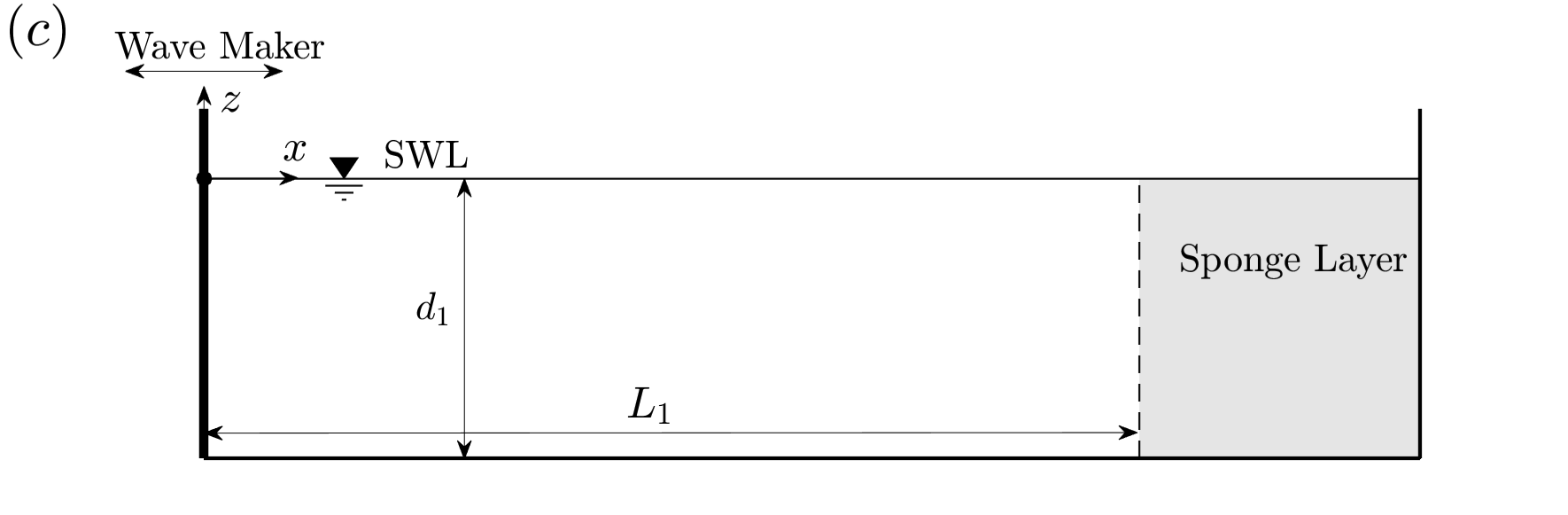}
\caption{Schematic of the side-view of the computational domain for the waves shoaling over $(a)$ a plane beach, $(b)$ an idealized bar, and $(c)$ a flat bed geometry.}
\label{fig1}
\end{figure}

Our numerical experiments are performed in a virtual wave tank with three different idealized bed geometries, illustrated in Figure~\ref{fig1}.  Cases include deep to shallow water transition conditions. 
We define the coordinate system $(x,y,z)$ such that $x$ and $y$ represent the along-tank and transverse directions respectively and $z$ is the vertical direction, positive upward and measured from the still water level. We note that waves are usually breaking over the bar crest or the down-wave slope for cases of shoaling over a bar  ($x>0$ in Figure~\ref{fig1}$b$). 

All model simulations are performed with the model initialized with quiescent conditions. In the LES/VOF model, we specify the total instantaneous free surface, $\eta_w$, and liquid velocity, $(u_w,w_w)$, at the model upstream boundary, $(x_w,y,z)$, for various incident wave conditions, including regular sinusoidal, focused packets, modulated wave trains, and irregular waves propagating over a flat bed or over a bar geometry, as well as regular cnoidal waves shoaling over a plane beach. 
In the BEM model, we specify solitary waves as initial condition on the free surface, using the elevation, potential and normal velocity derived from the \citet{tanaka1986stability} solution. Periodic cnoidal waves are specified at the upstream wavemaking boundary, similar to the LES/VOF model.
Table \ref{table1} summarizes the input parameters for all simulated cases. 

\begin{table}
  \begin{center}
{
\centering
\begin{tabular}{l ccccccccc}
     Case &$H_w\; (mm)$ 				&$T_w$ 	&$d_1$ &$L_1$ &$s$ &$\xi_0$ &$d_2$ &$L_2$  & $s_d$ \\[3pt]
     	      &	(or $S_g$)			&(s) 	&(m)   &(m)   &    &   (or $\Delta f / f_c$)     &(m)   &(m)       &  \\ [3pt]
\hline
     P1-r-LV	  &		80, 120, 180, 200, 240 		&4.0	&0.5  & 0     &5  & 3.9 - 2.3  &-   &-         &-  \\     
     P2-r-LV	  &	 150 		&4.0	&0.5  & 0     &10  & 1.43  &-   &-         &-  \\[3pt]     
     P3-r-LV	  &	 40, 150 		&4.0	&0.5  & 0     &20  & 1.38, 0.71  &-   &-         &-  \\[3pt]     
     P4-r-LV	  &	 90, 150, 200 		&4.0	&0.5  & 0     &40  & 0.46 - 0.31 &-   &-         &-  \\[3pt]     
     P5-r-LV	  &	 90, 150 		&4.0	&0.5  & 0     &100  & 0.18, 0.14  &-   &-         &-  \\[3pt]     
     P6-r-LV	  &	 90, 120, 150 		&4.0	&0.3  & 0     &200  & 0.09 - 0.07  &-   &-         &-  \\[3pt]     
     P7-s-LV	  &		240, 260, 270, 350, 500 		&-	&1.0  & 6.0     &8  & -  &-   &-         &-  \\[3pt]     
     P7-s-BM	  &		{240} 		&-	&1.0  & 20.0     &8  & -  &-   &-         &-  \\[3pt]     
     P8-s-BM	  &		300, 450, 600		&-	&1.0  & 20.0     &15  & -  &-   &-         &-  \\[3pt]  
     P9-s-BM	  &		200, 600		&-	&1.0  & 20.0     &100  & -  &-   &-         & - \\[3pt]     
     B1-r-LV	  &41, 43, 46, 46.2, 46.3, &1.01	&0.4   &6     &20  &0.30 - 0.25        &0.1   &2        &10 \\[3pt]     
			      & 46.5, 47, 50, 53, 59&	 	&   	&      &  	&        &	    &      &       \\[3pt]
     B2-r-LV	  &47, 50, 53, 59 &1.01	&0.4   &6     &100  &0.06 - 0.05        &0.1   &2        & 10\\[3pt]    
    B3-r-LV	  &24, 26, {26.5}, 27, 	&2.525	&0.4   &6     &20  &1.05 - 0.81        &0.1   &2       & 10 \\[3pt]     
    &27.5, 30, 34, 40&	 	&   	&      &  	&        &	    &      &       \\[3pt]
	B4-r-LV	  &26, 30, 30.5, 31	&2.525	&0.4   &6     &100  &0.21 - 0.16        &0.1   &2        & 10\\ [3pt]     
			  &32, 34, 40&	 	&   	&      &  	&        &	    &      &       \\[3pt]
     B5-f-LV	  &$(0.20, 0.21, 0.22,$	&$T_c : 1.14$	&0.6   &3     &20  & (0.75)       &0.2   &3        & 10 \\ [3pt] 
             	  &$ 0.23, 0.30)$	&	&   &     &  &        &   &        &  \\ [3pt]     
    B6-i-LV	  &$H_{rms}: 40$	&$T_p: 1.7$	&0.47   &0     &20  &    0.52    &0.12   &2        & 10 \\ [3pt]     
    B7-i-LV	  &$H_{rms}: 40$	&$T_p: 1.7$	&0.47   &0     &20  &    0.52    &0.17   &2        & 10\\ [3pt]  
    B8-s-BM	  &36, 40, 46, {46.6}&-	&0.4   &8     &20  &  -      &0.1   &2       &10 \\[3pt]     
    	  &47, 60, 80&	 	&   	&      &  	&        &	    &      &       \\[3pt]
    F1-f-LV	  &$(0.25, 0.3, 0.302, 0.31,$	&$T_c : 1.14$	&0.6   &16     &-  & (0.75)      &-   &-       & - \\ [3pt]     
        	  &$0.32, 0.42, 0.44, 0.46)$	&	&   &     &  &        &   &        &  \\ [3pt]     
    F2-f-LV	  &$(0.32, 0.36, 0.40)$	&$T_c : 1.33$	&0.6   &22     &-  & (1.0)       &-   &-       & - \\ [3pt]     
    F3-m-LV	  &$(0.160, 0.176)$	&$T_c : 0.68$	&0.55   &64     &-  & (0.0954)       &-   &-       & - \\ [3pt]     
\hline
\end{tabular}}
\end{center}
\caption{Input parameters for the simulated cases. Each case identifier has 3 parts indicating the geometry of the wave tank (P: planar beach, B: barred beach, F: flat bed; numbers: various geometry parameters), the type of the incident waves (r: regular, i: irregular, s: solitary waves, f: focused packets, m: modulated wave trains), and the numerical model (LV: LES/VOF, BM: FNPF-BEM) respectively. Here, $H_w$ and $T_w$ are the wave height and period of the regular waves at the wavemaker, and $\xi_0 = s^{-1}/\sqrt{H_0/L_0}$ is the surf-similarity parameter \citep{battjes-icce74}; the rest of the variables are defined in Figure \ref{fig2}.}\label{table1}
\end{table}

\subsubsection{Focused wave packets}\label{sec3.2.1}

The input focused wave packet was composed of $N=32$ sinusoidal components of steepness $a_nk_n, n=1,\cdots,N$, where $a_n$ and $k_n$ are the amplitude and wave number of the $n$th frequency component.  The steepness of individual wave components is taken to be constant across the spectrum, or  $a_1k_1 = a_ik_i = ...=a_N k_N = S_g/N$ with $S_g = \sum _{n=1}^{N} a_n k_n$ taken to be a measure of the wave train global steepness. Based on linear theory, the free surface elevation at the wavemaker for 2D wave packets focusing at $x=x_f$ is given by \citep{Rapp-Melville:1990,Derakhti-Kirby:2014}
\begin{equation}
\eta_w = \sum_{n=1}^{N}a_n\cos[2\pi f_n(t-t_f)+{k_n(x_f-x_w)}] \label{eq7}
\end{equation}
where $f_n$ is the frequency of the $n$th component, $x_f$ and $t_f$ are the predefined, linear theory estimates of location and time of the focal point respectively.  The discrete frequencies $f_n$ are uniformly spaced over the band $\Delta f = f_N - f_1$ with the central frequency defined by $f_c = {1}/{2}(f_N + f_1)$. 

\subsubsection{Modulated wave trains}\label{sec3.2.2}

For cases of modulated wave trains, we use the bimodal wave approach of \citet{Banner-Peirson:2007}, with  free surface elevation at the wavemaker given by 
\begin{equation}
\eta_w = a_1 \cos(\omega_1t) + a_2 \cos(\omega_2t -\frac{\pi}{18}), \label{eq8}
\end{equation}
where $\omega_1=2\pi f_1$, $\omega_2=\omega_1+2\pi\Delta f$, $S_g = a_1k_1+k_2a_2$ and $a_2/a_1 = 0.3$.  Increasing the global steepness $S_g$ increases the strength of the resulting breaking event in both focused packets and modulated wave trains. 

\subsubsection{Irregular wave trains}\label{sec3.2.3}

For irregular wave cases, $\eta_w$ is prescribed using the first $N=2500$ Fourier components of the measured free surface time series at the most offshore gauge of the cases experimentally studied by \citet{Mase-Kirby:1992} with $T_p = 1.7$s, given by
\begin{equation}
\eta_w = \Sigma_{n=1}^N a_n \cos(\omega_n t + \epsilon_n) \label{eq9}
\end{equation}
where $a_n$ and $\epsilon_n$ are the amplitude and phase of the $n$th Fourier component based on the measured free surface time series, and $\omega_n$ is the angular frequency of the $n$th Fourier component. \citet{Mase-Kirby:1992} specified wavemaker conditions for irregular waves based on a Pierson-Moskowitz spectrum.  Waves then propagated shoreward over a sloping planar beach. Here the same incident waves are used but shoal over an idealized bar. Liquid velocities for each spectral component are calculated using linear theory and then superimposed linearly at the wavemaker.  No correction for second order effects was made.

\subsubsection{Regular weakly dispersive, nonlinear waves}\label{sec3.2.4}

For cnoidal waves, we use the theoretical relations for $\eta_w$ and $(u_w,w_w)$ as given in \citet{Wiegel:1960}.
Initial conditions for solitary wave tests were specified using the solution for finite amplitude waves due to \citet{tanaka1986stability}.  This initial condition represents a very accurate numerical solution to the full Euler equations, and is more suitable for use here with the fully nonlinear numerical codes being used than the standard first-order Boussinesq solitary wave solution \citep[{\it e.g.}][]{grilli1996numerical}.

\subsection{Definition of local geometric parameters used in the analysis} \label{sec3.3}

\begin{figure}
\centering
\includegraphics[width=0.9\textwidth]{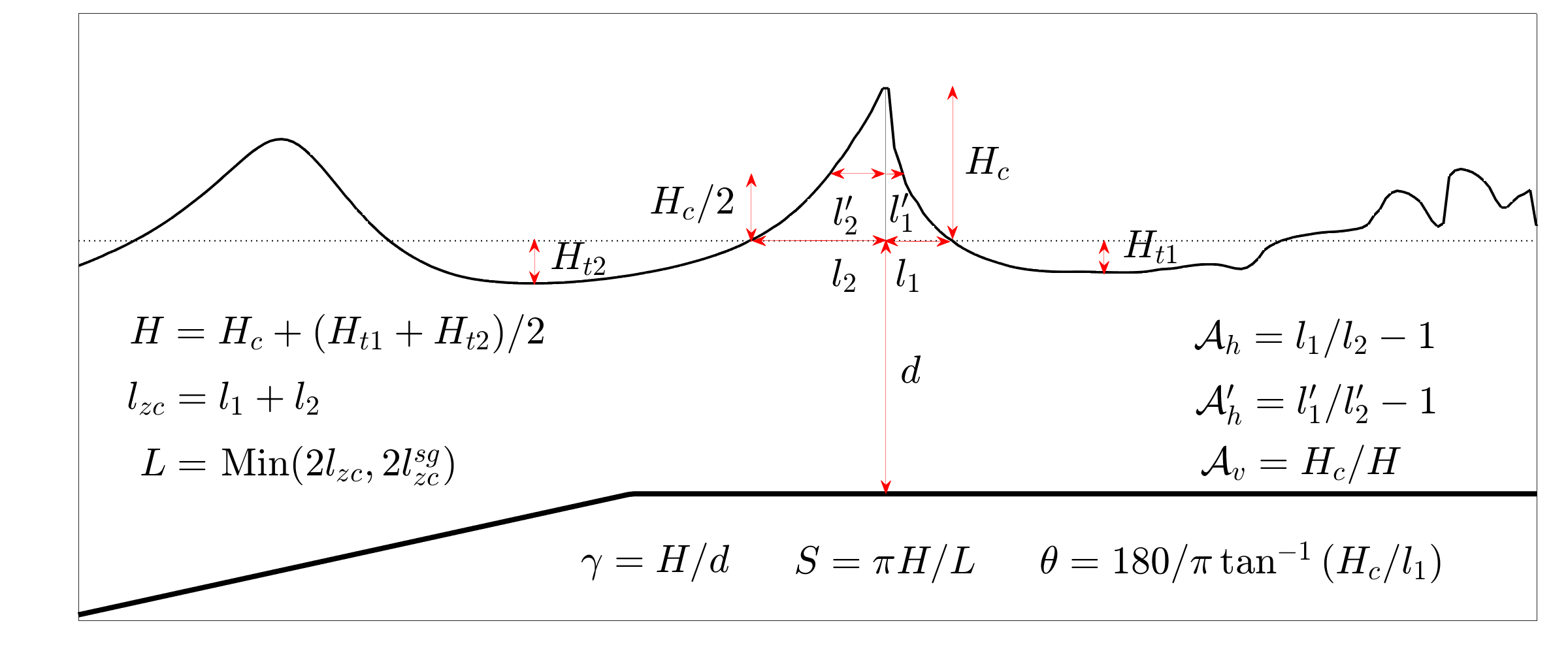}
\caption{Local geometric parameters describing an evolving wave crest. Here $l_{zc}^{sg}$ is defined in Appendix B and represents a length scale obtained from a skewed-Gaussian fit to the crest region. Dotted and thick solid lines show the still water and the bed elevations respectively. The incident waves are propagating from left to right.}
\label{fig2}
\end{figure}

Definitions of the various local geometric parameters for an evolving wave crest are described in Figure~\ref{fig2}. Among these, the height $H$ and length $L$ of the carrier wave need to be defined first. We define the local wave height $H$ as the sum of a crest elevation and averaged trough elevations before and after the crest, $H = H_c + (H_{t1}+H_{t2})/2$. Following \citet{Derakhti-Kirby:2016}, \citet{Derakhti-etal:2018} and \citet{tian-etal-pf08}, we define the local wave length $L = 2l_{zc}$, where $l_{zc} = l_1+l_2$ (Figure~\ref{fig2}) is the distance between the two consecutive zero-crossing points adjacent to the crest. We note that the zero-crossing point on the back face of the wave may have noticeably large fluctuations due to the presence of higher harmonics in shallow water cases (Figure~\ref{figB1}$b$) or high-frequency components in random waves, etc. Further, in some shallow water cases, {\it e.g.}, solitary waves, there are no zero-crossing points and thus $l_{zc}$ can not be defined. To resolve these issues, as explained in Appendix B, we fit a skewed-Gaussian function to the instantaneous wave profile and then estimate a length scale $l_{zc}^{sg}$ from the skewed-Gaussian fitting (Figure~\ref{figB1}). Finally, we take $L =$ Min$(2l_{zc},2l_{zc}^{sg})$ as the local wave length. As discussed in Appendix B, using definitions other than those used here may vary the estimated $H$ values for extreme waves by up to $10\%$. However, the sensitivity of the estimated $L$ values at breaking onset are noticeably larger, especially for shallow cases.

Based on the local definitions of $H$ and $L$, we define nonlinearity parameters
\begin{equation}
\gamma = H/d,
\end{equation}
the height to depth ratio and 
\begin{equation}
S=\pi H / L \label{eq11}
\end{equation}
the local steepness.  Following \citet{Beji:1995}, we also define a wave Froude number
\begin{equation}
F = gH/2c_{lin}^2,\label{eq12}
\end{equation}
where $c_{lin}^2 = gk^{-1}\tanh{kd}$ and $k = 2\pi/L$ is the local wave number. We note that $F$ simplifies to $\gamma/2$ in shallow water and to $S$ in deep water.  $F$ may thus be considered to be a unified nonlinearity parameter in arbitrary depth \citep{Beji:1995,Kirby:1998}. Further, using the results from linear theory, we can readily obtain $F = u_{lin}/c_{lin}$, 
where $u_{lin}$ is the linear theory prediction of the particle velocity at the horizontal crest position and at the mean water level. 
All of these properties suggest that $F$ is a preferable diagnostic geometric parameter compared to $\gamma$ and $S$ for a unified breaking onset criterion in arbitrary depth.  

We define a wave front slope $\theta$ in degrees by
\begin{equation}
    \theta= \frac{180}{\pi} \tan^{-1}(H_c/l_1),\label{eq13}
\end{equation}
where $H_c/l_1$ is the crest front steepness (see Figure~\ref{fig2}). 
We further define $\mathcal{A}_v = H_c/H$ and $\mathcal{A}_h = l_1/l_2-1$, which represent instantaneous vertical and horizontal asymmetry of the evolving crest, and are related to the statistical third-order moments, normalized wave skewness $\overline{\eta^3}\,/\,\overline{\eta^2}^{3/2}$ and asymmetry $\overline{\mathcal{H}(\eta)^3}\,/\,\overline{\eta^2}^{3/2}$ (where $\mathcal{H}$ denotes the Hilbert transform), respectively. Finally, we define $\mathcal{A}^{\prime}_h = l^{\prime}_1/l^\prime_2-1$, which represents the horizontal asymmetry of the shape of the crest but only considering the upper half part of the crest. $\mathcal{A}^{\prime}_h$ is also applicable for crests without zero-crossing points and is a more robust measure compared to $\mathcal{A}_h$ for crests with noticeable irregularity at their back face (Figure~\ref{figB1}$b$). The parameter $\theta$ is often used as the diagnostic criterion for the onset of breaking in Boussinesq models using eddy viscosity-type dissipation to model breaking \citep[see, for example,][]{schaffer-etal-ceng93, kennedy-etal-ww00a}.

\section{Results}

In this section, we examine the occurrence of breaking onset on basis of the parameter $B$ of \citet{Barthelemy-etal:2018} for representative breaking and non-breaking incident waves in intermediate depth and shallow water. The results for steepness-limited wave breaking cases for both focused packets and modulated wave trains \citep{Derakhti-etal:2018} are also presented. 
In \S5, we show that several geometric criteria for predicting the onset of breaking are not uniformly robust, although a measure of surface slope spanning the crest to trough front face of the wave, such as $\theta$ (Eq.~\ref{eq13}), in combination with a wave Froude number $F$ (Eq.~\ref{eq12})  is seen to be remarkably accurate.

The local energy flux parameter $\boldsymbol{B}$ introduced by \citet{Barthelemy-etal:2018} is defined at the wave crest region as
\begin{equation}
    \boldsymbol{B} = \boldsymbol{\mathcal{F}}/E|\boldsymbol{C}| 
    \label{eq140}
\end{equation}
where $\boldsymbol{\mathcal{F}} = \boldsymbol{U}(p+E)$ is the local flux of mechanical energy/unit volume, $E$ is the mechanical energy/unit volume, and $\boldsymbol{U}$ is the local liquid velocity. The wave crest translates with propagation speed $C = |\boldsymbol{C}|$, which is generally time-dependent. On the free surface, the pressure $p$ is taken to be zero, reducing the expression for $\boldsymbol{B}$ to  
\begin{equation}
    {B} = U/C 
    \label{eq14}
\end{equation}
where $U$ is the ensemble of liquid velocity at the wave crest in the direction of wave propagation. Although Equation (\ref{eq14}) appears similar to the kinematic breaking onset criterion \citep[][\S3.2]{Perlin-etal:2013}, it represents the normalized flux of mechanical energy at the crest, and then should be considered as a dynamical criterion. 

\begin{figure}
\centering
\includegraphics[width=\textwidth]{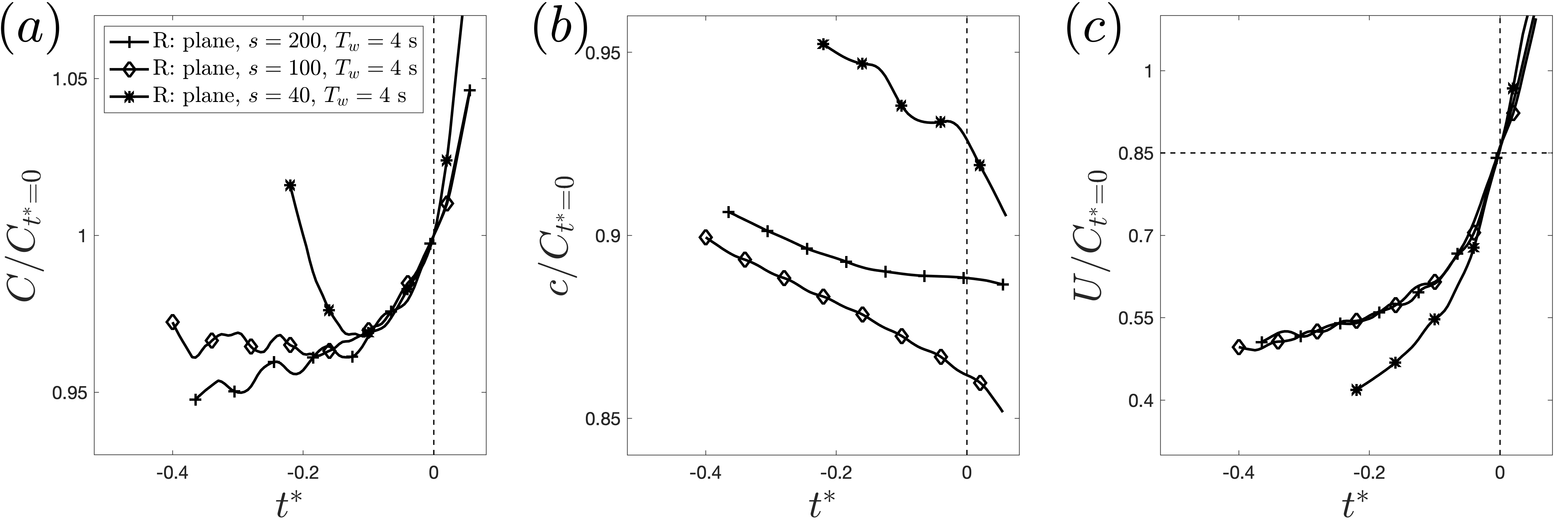}
\includegraphics[width=\textwidth]{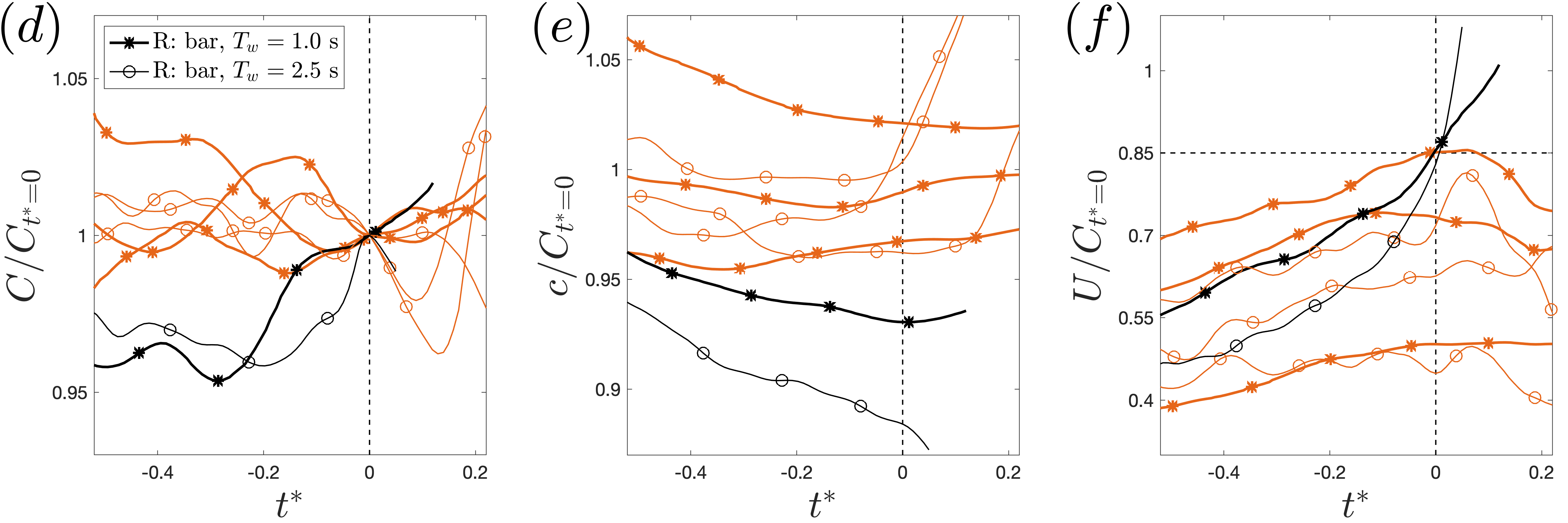}
\includegraphics[width=\textwidth]{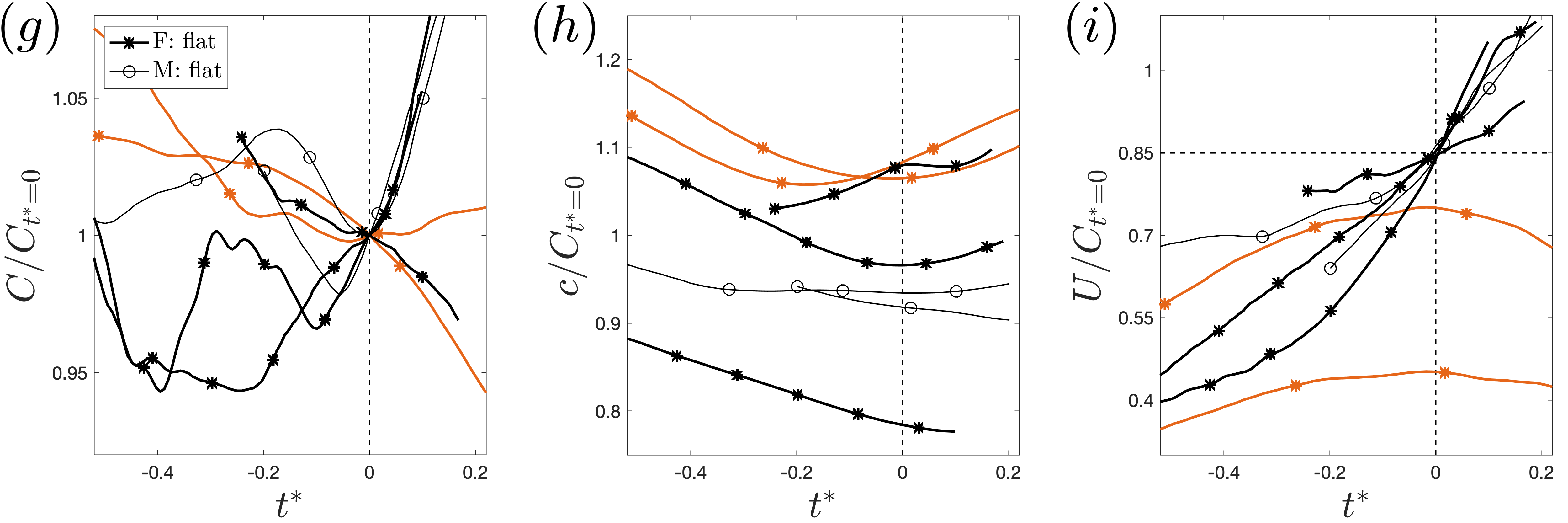}
\caption{Examples of the temporal variation of $(a,d,g)$ the crest propagation speed $C$, $(b,e,h)$ phase speed $c = \sqrt{gk^{-1}\tanh{[k(d+H_c)]}}$, and $(c,f,i)$ the horizontal particle velocity at the crest $U$, all normalized by the corresponding $C$ value at $t^*=0$, for breaking (black symbols and lines) and non-breaking (orange symbols and lines) crests in $(a,b,c)$: regular waves (R) shoaling over a plane beach with slope $m = 1/s$, $(d,e,f)$: regular waves (R) propagating over a bar, and $(g,h,i)$: focused packets (F) and modulated wave trains (M) in deep and intermediate water.}
\label{fig5}
\end{figure}

Frames (a), (d), and (g) of Figure~\ref{fig5} show examples of the computed temporal variation of $C$ from shallow to deep water, in which values of $C$ are normalized by their corresponding values at the time $t^*=0$ marking either breaking onset or, for non-breaking cases,  the occurrence of maximum crest elevation. In this and subsequent plots, the color black for curves or points indicates cases where breaking occurs, while orange indicates non-breaking cases.  Frames (b), (e), and (h) show results for an estimate of phase speed $c$ based on an approximate nonlinear dispersion relation 
$c = \sqrt{gk^{-1}\tanh{[k(d+H_c)]}}$ which is slightly different than that proposed by \citet{Booij:1981} (replacing $H/2$ by $H_c$). The behavior of crest translation speed $C$ is seen to be distinctly different from estimates based on dispersion relations for regular waves. The results show that the ratio $c/C$ around $t^*=0$ ranges between 0.8 and 1.1 in most cases. 

We note that $C$ is obtained by calculating the rate of change of the horizontal location of an evolving crest, e.g., $x_{\eta_c}$ if the crest is propagating in the $x$ direction. In both BEM and LES/VOF frameworks, $x_{\eta_c}$ may occur between the grid locations, and thus a local fitting (or smoothing) to the predicted free surface locations ($\eta(x,y,t)$) is needed to obtain a robust estimate of $x_{\eta_c}$ for each evolving crest. Such local fitting (or smoothing) also removes the potential noise in the calculated $C$ values due to the existence of local maxima in the crest region due to the presence of relatively high frequency waves, especially when they are propagating in the direction opposite to that of the dominant wave. 
Although implementing local fitting (or smoothing) for predicted maximum $\eta$ values and their locations significantly improves the estimation of $C$, in some cases there are still some small undulations in the $C$ values (as shown in the left column of Figure~\ref{fig5}) obtained from tracking the location of $\eta_c$ (e.g., $C = dx_{\eta_c}/dt$). In deep water cases, a part of the observed variation in $C$ is due to the interaction of the carrier wave with a wave packet \citep{Banner-etal:2014}. 

In addition, the estimation of $C$ will be challenging in cases in which the crest region is relatively flat. One clear example of such cases is the time at which an evolving crest reaches the shoreline and the wave rapidly surges the up slope without overturning; such cases are detailed later in the text. 
Considering these uncertainties we can write $C = C_e\pm\Delta C$ where $C_e$ is the exact propagation speed of the evolving crest and $\Delta C$ represents the corresponding uncertainty estimate. The results suggest that $\Delta C/C_e<0.01$ prior to $t^*=0$ in the simulated cases in which the crest region has a resolved curvature in the considered discretization.    

In the BEM model, $U$ is the actual particle velocity on the free surface at the wave crest. In the LES/VOF model, we set $U$ as the maximum of the computed horizontal near-surface velocity over the computational cells in the range $x_{\eta_c}\pm3\Delta x$. We also perform a simple smoothing, using the moving average method, on the $U$ time series for each evolving crest before calculating $B$ values.    
Frames (c), (f), and (i) of Figure~\ref{fig5} show examples of the temporal evolution of $U$ normalized by their corresponding $C$ values at the time $t^*=0$, $C_{t^*=0}$. In Figure~\ref{fig5}, all $C$, $c$, and $U$ values that are correspond to an evolving crest are normalized by a single value $C_{t^*=0}$, the propagation speed of the crest at the time $t^*=0$. Thus $U/C_{t^*=0}$ is not equal to $B$ for $t^*\ne 0$.
Our results show that $U$ significantly increases as an evolving crest approaches the break point $t^*=0$, as opposed to $C$, which varies by less than $5\%$ in the range $-0.4<t^*<0$ for cases of shoaling over gentle to moderate slopes or cases in deep and intermediate depth water. For these cases, the results suggest that the variation in $B$ is mainly related to variation in $U$.

We also write $U$ in terms of the exact value ($U_e$) and an uncertainty estimate ($\Delta U$), $U= U_e\pm\Delta U$, in which the results indicate that $\Delta U/U_e < \Delta C/C_e$ for most cases. 
Thus, we can write
\begin{equation}
    B = \frac{U_e\pm\Delta U}{C_e \pm \Delta C} = \frac{U_e}{C_e}\times \frac{1\pm\Delta U/U_e}{1\pm\Delta C/C_e}=B_e(1\pm\Delta U/U_e){\Big(}1\pm\Delta C/2C_e+O([\Delta C/C_e]^2)\Big).\label{Berror1}
\end{equation}
where $B_e$ represents the exact value of $B$, and then the uncertainty in the estimated $B$ values, denoted by $\Delta B$, reads in relative value as
\begin{equation}
\pm\Delta B/B_e = \pm\Delta U/U_e\pm\Delta C/2C_e + O([\Delta C/C_e]^2,[\Delta C/C_e][\Delta U/U_e]).\label{Berror2}
\end{equation}
\noindent Based on these results and taking $\Delta U/U_e <\Delta C/C_e<0.01$, the uncertainty in the estimated $B$ values from our numerical experiments (describe below) will be $\Delta B/B_e<0.015$ for the cases in which the crest region has a resolved curvature in the considered discretization. In particular, $\Delta B< 0.013$ for $B$ values approaching the threshold breaking-precursor value $B_{th}$, which varied in the range $[0.85,0.86]$ in the numerical cases of \citet{Barthelemy-etal:2018}. Further, \citet{Saket-etal:2017,Saket-etal:2018} reported an uncertainty estimation of $\Delta B = 0.020$ for their experimental
measurements.

\begin{figure}
\centering
\includegraphics[width=\textwidth]{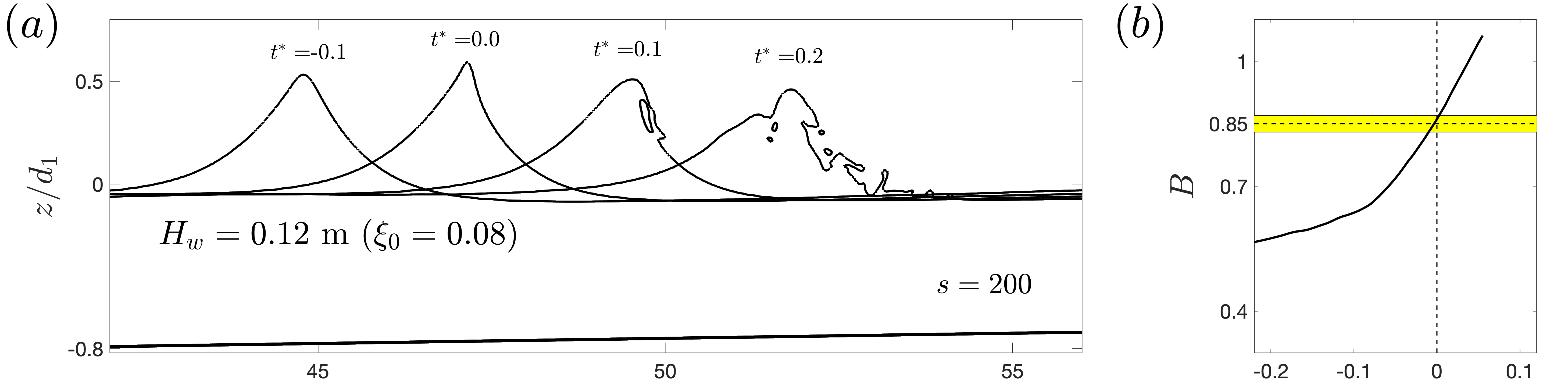}
\includegraphics[width=\textwidth]{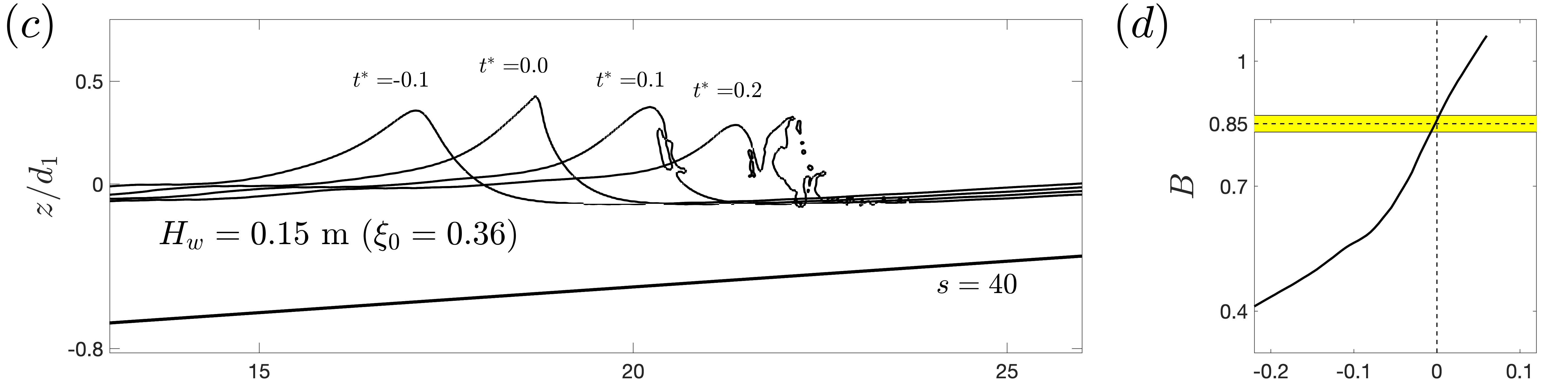}
\includegraphics[width=\textwidth]{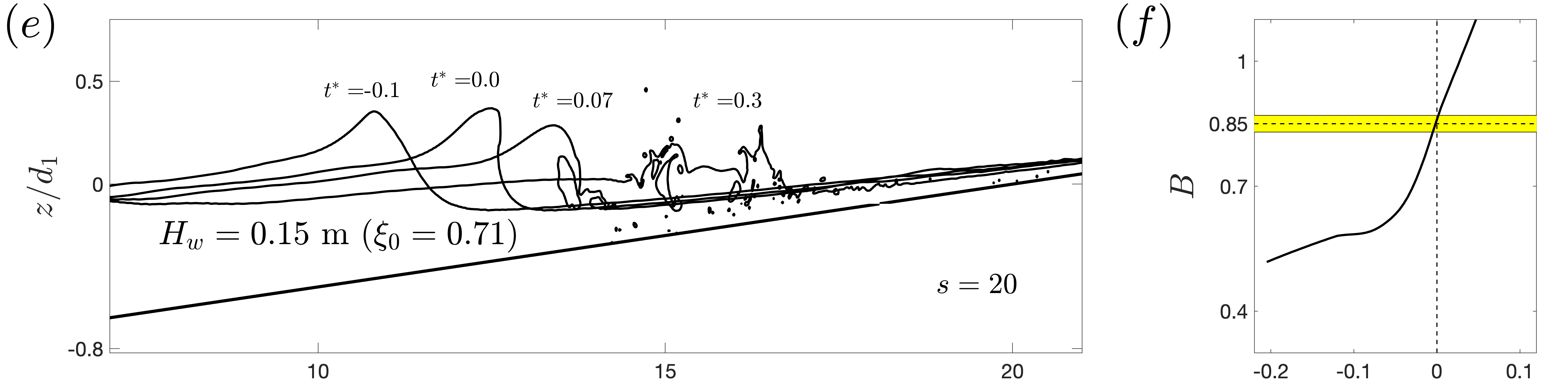}
\includegraphics[width=\textwidth]{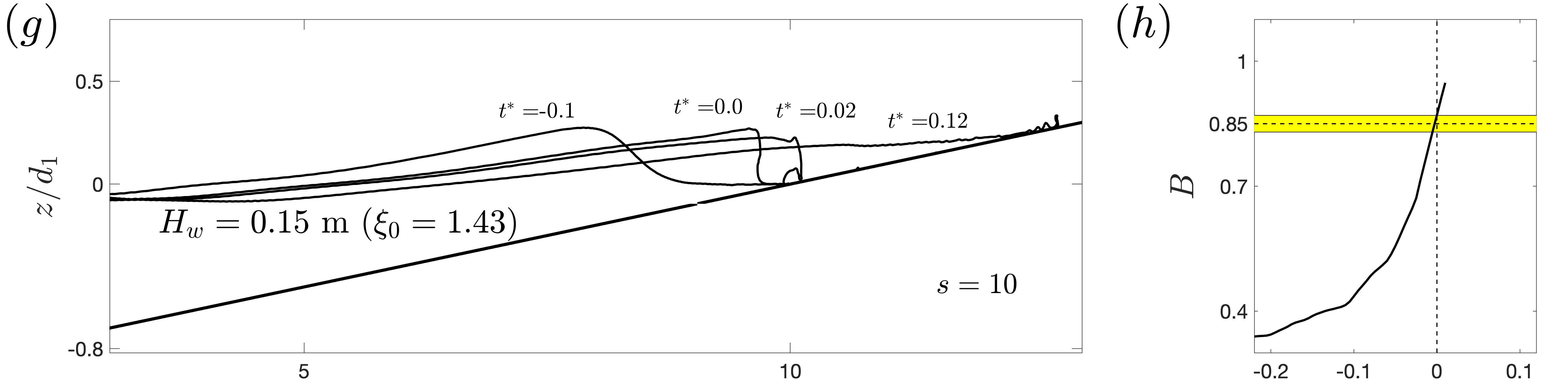}
\includegraphics[width=\textwidth]{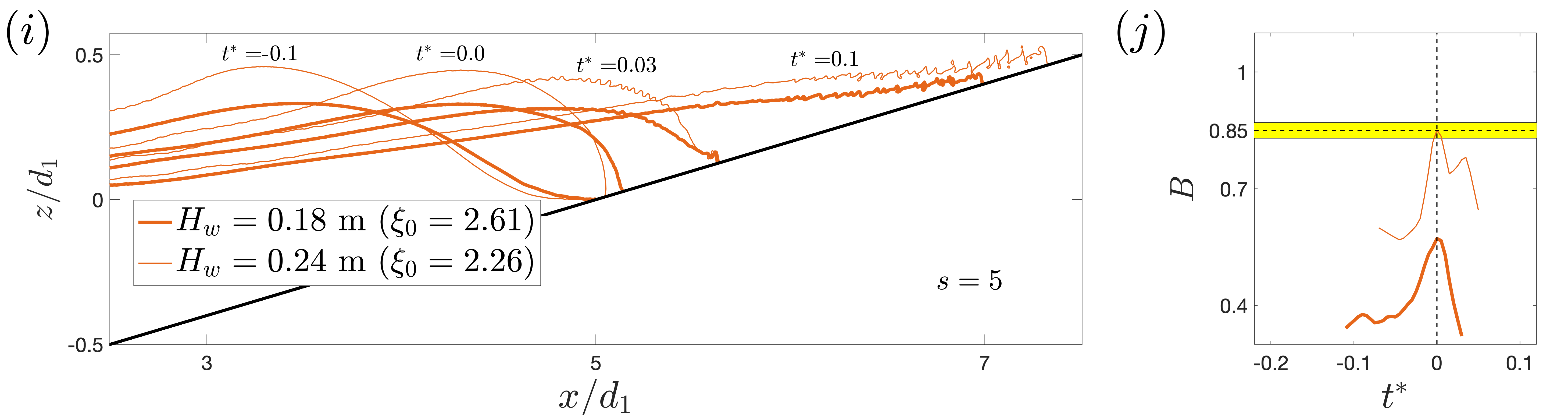}
\caption{$(a,c,e,g,i)$: Snapshots of free surface elevations and $(b,d,f,h,j)$: temporal evolution of the breaking onset parameter $B$ for regular waves ($T_w = 4$ s) propagating over a plane beach with a slope $m = 1/s$, demonstrating a transition from spilling to collapsing and surging breaking with an increasing $\xi_0 = s^{-1}/\sqrt{H_0/L_0}$ (see Table~\ref{table1}). Here $d_1$ is the still water depth at the beginning of the plane slope segment (Figure~\ref{fig1}$a$). Cases without an apparent overturning crest are indicated in orange. All results are obtained using the LES/VOF model.}
\label{fig6}
\end{figure}

\subsection{Results for waves shoaling over a plane beach}

Figure~\ref{fig6} shows examples of the evolution of regular waves (with $T_w = 4$ s) over a plane beach with a slope $m = 1/s$; including shoaling, breaking onset and progression of breaking crests; and the corresponding temporal variation of the breaking onset parameter $B$ for the tracked crests. The incident waves cover a wide range of $\xi_0$ values, demonstrating a transition from spilling breaking, frames (a-b), to collapsing and surging breaking, frames (g-j), events. 
We observe that $B$ always passes $B_{th}\sim 0.85$ prior to breaking onset in cases in which breaking is due to the initiation of instability in the crest region (i.e., spilling or plunging breakers). In these cases, we also observe that $B$ eventually passes 1 shortly after the precursor value $B_{th}=0.85$ is reached, and the time scale $\Delta t_{onset} = t_{B = 1}-t_{B=B_{th}}$ is a decreasing function of $\xi_0$ for breaking waves with the same wave period ($T_w$ here).
Note that in shorebreaks, $\Delta t_{onset}$ is relatively small and estimating $B$ is challenging due to rapid changes, uncertainty, and ambiguity in defining $x_{\eta_c}$ after $B\sim B_{th}$ is passed.
For example, in the shorebreak case shown in frames (g-h), the calculation of $B$ is terminated before reaching $B=1$ due to a poor estimation of $C$ as the wave transitions through the time at which $B = B_{th}$. Frames (i) and (j) show the results for two cases with $\xi_0 = 2.26$ and 2.61 surging over a slope of 1/5 or slope angle $11.31^{\circ}$. In both cases, the initiation of instability occurs at the toe of the wave, and the maximum $B$ values remain below $B_{th}$. The slope $1/5$ is thus close to the maximum slope with potential breaking for the considered waves.

\begin{figure}
\centering
\includegraphics[width=\textwidth]{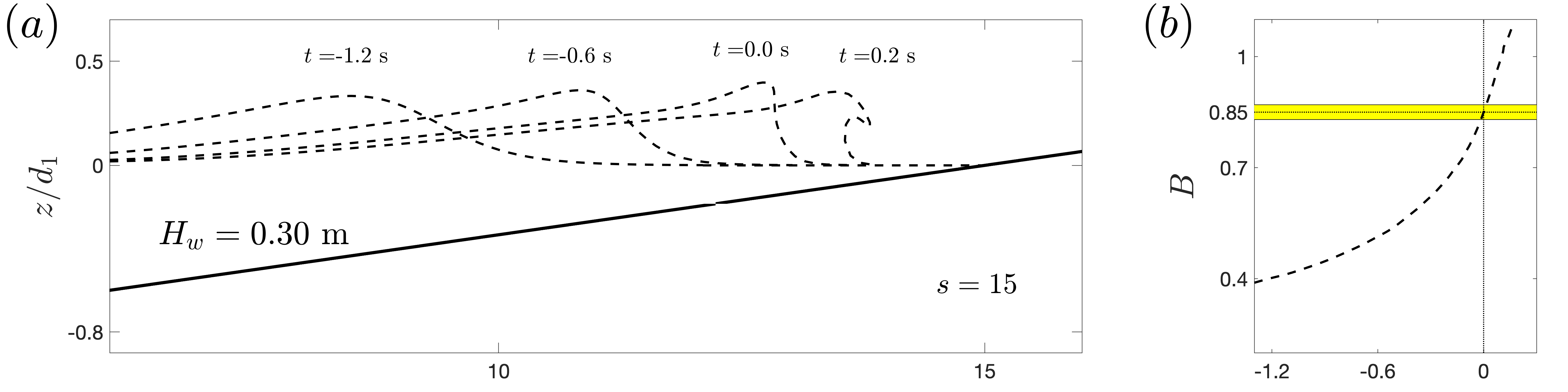}
\includegraphics[width=\textwidth]{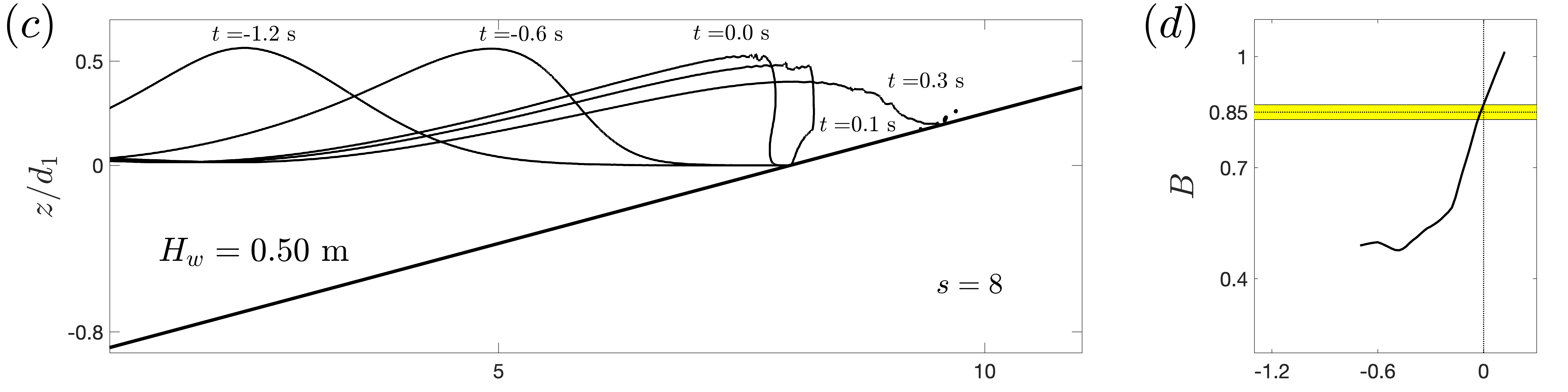}
\includegraphics[width=\textwidth]{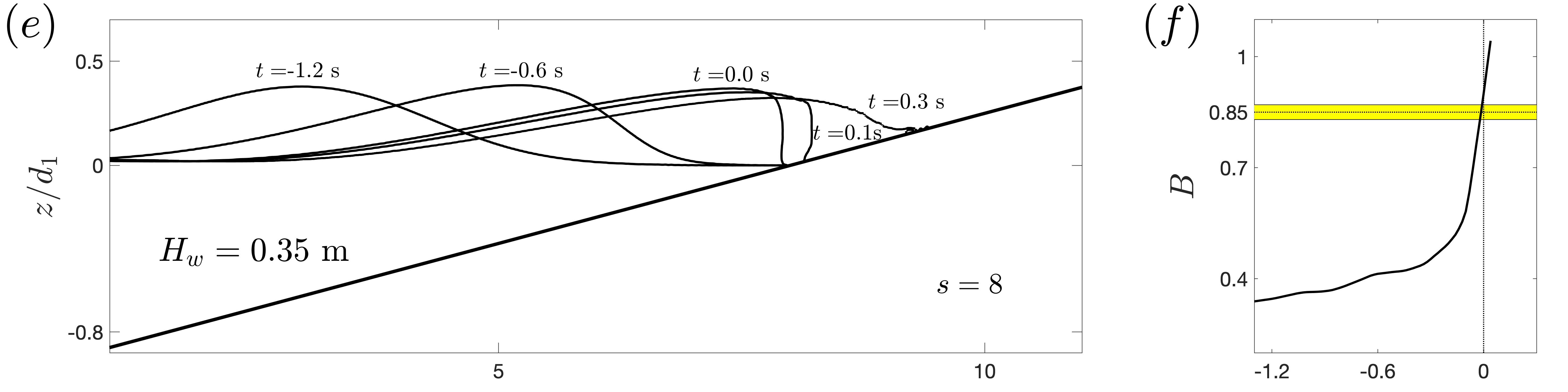}
\includegraphics[width=\textwidth]{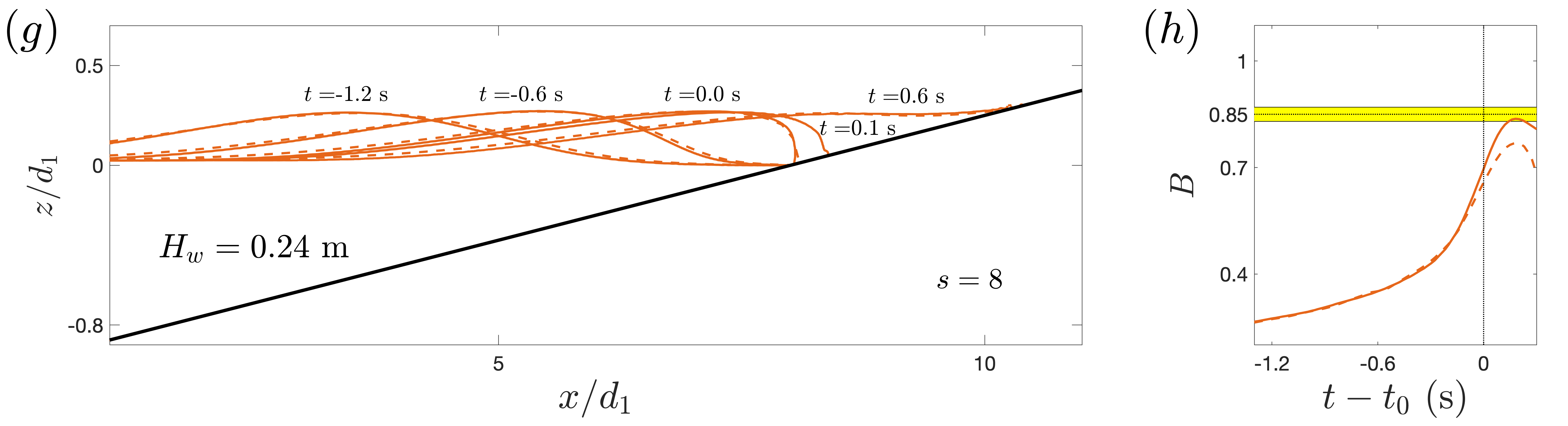}
\caption{$(a,c,e,g)$: Snapshots of the free surface elevations and $(b,d,f,h)$: the temporal evolution of the breaking onset parameter $B$ for solitary waves propagating over a plane beach with a slope $m = 1/s$. Dashed and solid lines represent the results for cases simulated using the BEM and LES/VOF models respectively.}
\label{fig7}
\end{figure}

Figure~\ref{fig7} shows similar results to those of  Figure~\ref{fig6} for solitary wave cases shoaling over steep beaches simulated using the BEM (dashed lines) and LES/VOF (solid lines) models. Here and subsequently, dashed and solid lines represent results of simulations using the BEM and LES/VOF models, respectively. 
Breaking of solitary waves on plane slopes from 1/100 to 1/8 was studied using the BEM model by \citet{grilli1997breaking}, who reported no breaking for slopes greater than $12^{\circ}$. Using a least-square error method based on their numerical experiments, \citet{grilli1997breaking} proposed a maximum limit for non-breaking solitary waves shoaling on a slope $m = 1/s$ given by $H^{m}_w = 16.9d_1/s^2$. They also introduced a parameter $\zeta_0 = 1.521/s\sqrt{H_w/d_1}$ and characterized the type of their breaking cases based on $\zeta_0$ as surging when $0.30<\zeta_0<0.37$, plunging when $0.025<\zeta_0<0.30$, and spilling when $\zeta_0<0.025$. 

Figure~\ref{fig7}a shows the BEM model results for the evolution of a plunging breaking solitary wave on a slope 1/15 with $H_w = 0.30$ m $> H^{m}_w =0.08$ m and $\zeta_0 = 0.19$ ($d_1 = 1$ m). Frames (c) and (e) show results of the LES/VOF model for two cases on a slope 1/8 ($H_w^m = 0.264$) with $H_w = 0.50$ m ($\zeta_0 = 0.27$) and $H_w = 0.35$ ($\zeta_0 = 0.32$). For all three cases shown in frames (a-f), the occurrence and breaking type of the incident solitary waves predicted by both the BEM and LES/VOF models are consistent with the predictions from $H^{m}_w$ and $\zeta_0$ \citep{grilli1997breaking}. In all three cases, we observe that the corresponding $B$ parameter reaches $0.85$ close to a time at which a vertical tangent appears on the crest front face. As in Figure~\ref{fig6}, we also observe that $B$ eventually passes 1 shortly after $B_{th}$ is reached for all breaking solitary waves, and the time scale $\Delta t_{onset} = t_{B = 1}-t_{B=B_{th}}$ is a decreasing function of $\zeta_0$, consistent with the trend observed for regular waves with respect to $\xi_0$ (Figure~\ref{fig6}).

\begin{figure}
\centering
\includegraphics[width=\textwidth]{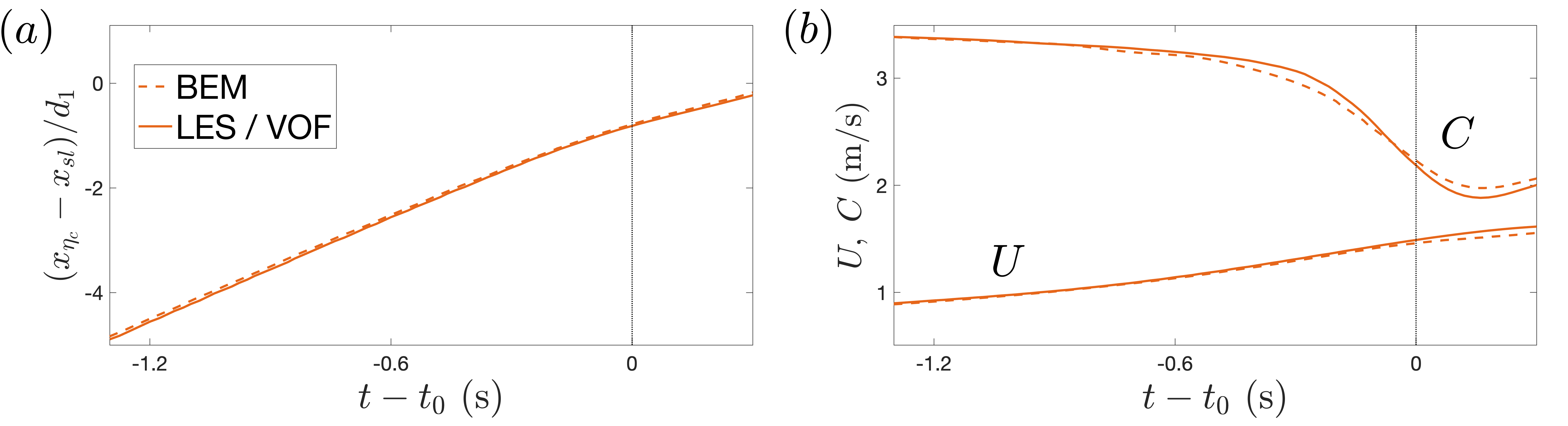}
\caption{Temporal evolution of $(a)$ $x_{\eta_c}$ the horizontal location of the crest, and $(b)$ $C$ crest propagation speed and $U$ particle horizontal velocity at the crest for the non-breaking solitary wave shown in Figure~\ref{fig7}g. Dashed and solid lines show results of simulations cases with the BEM and LES/VOF models, respectively. Here $x_{sl}$ is the cross-shore location of the shoreline. }
\label{fig8}
\end{figure}

Figure~\ref{fig7}g shows the evolution of a non-breaking solitary wave on a slope 1/8 with $H_w = 0.24$ m $< H^{m}_w =0.264$ m predicted by both the BEM and LES/VOF models. Frame (h) shows the calculated $B$ curves from both model results. In this case, $t_0$ represents the time of occurrence of maximum crest elevation, as opposed to Frames (a-f) in which $t_0$ represents the time when $B \approx 0.85$. The maximum $B$ values, $B_m$, from both models remain below $B_{th}$; however, $B_m$ calculated from the BEM model, is approximately $6\%$ smaller than that from the LES/VOF model results. Figure~\ref{fig8} shows the temporal variations of $x_{\eta_c}$, $C$ and $U$ predicted by both models. The maximum difference between $C$ and $U$ is approximately $4\%$. Before the time at which the crest maximum is reached ($t<t_0$), $U$ from the BEM model is almost the same as that predicted by the LES/VOF model except close to the crest maximum time, when the difference between the two predictions reaches $1\%$.  The BEM model prediction for $C$ is smaller and greater than that predicted by the LES/VOF model for $t<t_0$ and $t>t_0$, indicating that the BEM-predicted wave crest is pitching forward somewhat more slowly than the LES/VOF-predicted crest. The discrepancy in the corresponding $B$ values is a maximum after $t>t_0$, with a value of $\approx6\%$.   
The discrepancy between the BEM and LES/VOF results is partly due to their different spatial resolution ($\Delta x_{BEM} \approx 13 \Delta x_{LES/VOF}$) and the neglect of the bed friction and viscous effects in the BEM model. Further, a part of the discrepancy is related to the uncertainty in the estimation of $C$ as the crest region becomes relatively flat, particularly for surging/shorebreak cases. Overall, we find that the two modeling approaches provide consistent estimates of liquid velocity and crest geometry evolution in cases where adequate spatial and temporal resolutions are used. This conclusion is further supported by the general consistency observed between intermediate and deep water results in the studies of \citet{Barthelemy-etal:2018} and \citet{Derakhti-etal:2018}, and contrasts with the negative evaluation of the LES/VOF approach made in \citet{pizzo-melville-jfm19}.  This is further supported by validations presented in the Appendix.

\subsection{Results for waves shoaling over an idealized bar}

As mentioned in the introduction section, $B_{th}$ may be considered as a precursor or indicator of breaking onset, with any wave for which $B$ becomes greater than $B_{th}$ inevitably breaking a short time ($\Delta t_{onset}$) later. However, one still needs to closely examine the behavior of $B$ in the transition from breaking to non-breaking cases in shallow water, including marginal breaking events, to confirm the validity of $B_{th}$ in a universal sense. 
The transition from breaking to non-breaking of shoaling waves over a plane beach may only occur close to the shoreline, where an accurate estimation of $B$ is challenging, as discussed above. Thus we consider the behaviour of $B$ for regular, irregular, and solitary waves, as well as focused packets, propagating over a submerged bar (Figure~\ref{fig1}b), with an emphasis on marginally breaking cases. In the following, we present and discuss the computed temporal variation of $B$ for cases of simulated regular and solitary waves. Cases with irregular waves and focused wave packets will be reported elsewhere.   

\begin{figure}
\centering
\includegraphics[width=\textwidth]{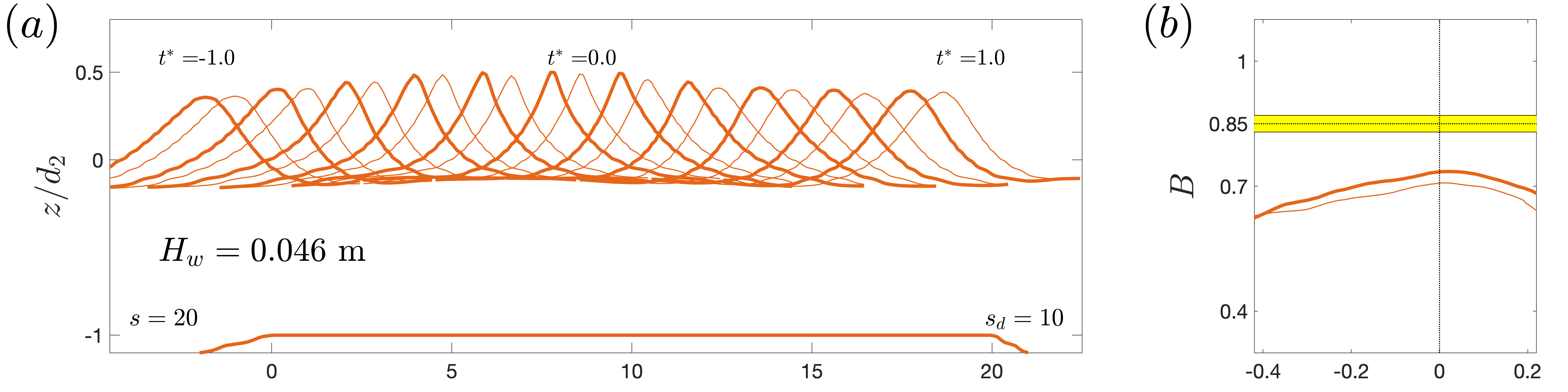}
\includegraphics[width=\textwidth]{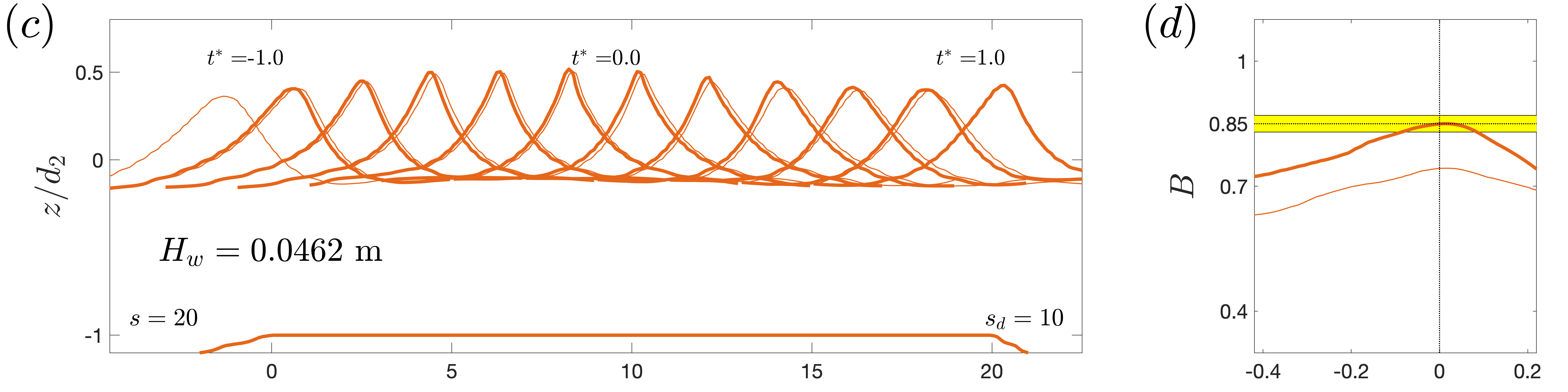}
\includegraphics[width=\textwidth]{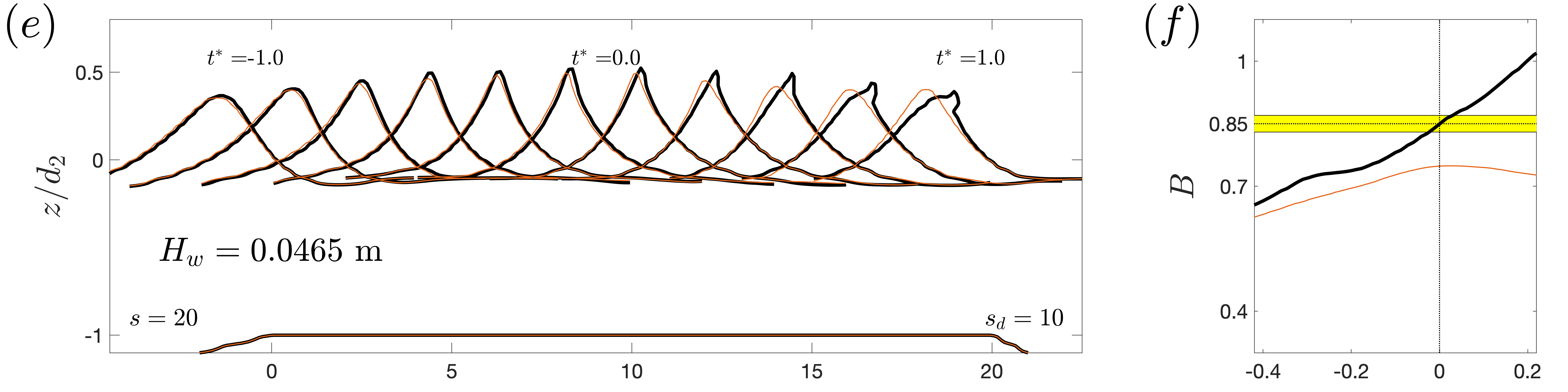}
\includegraphics[width=\textwidth]{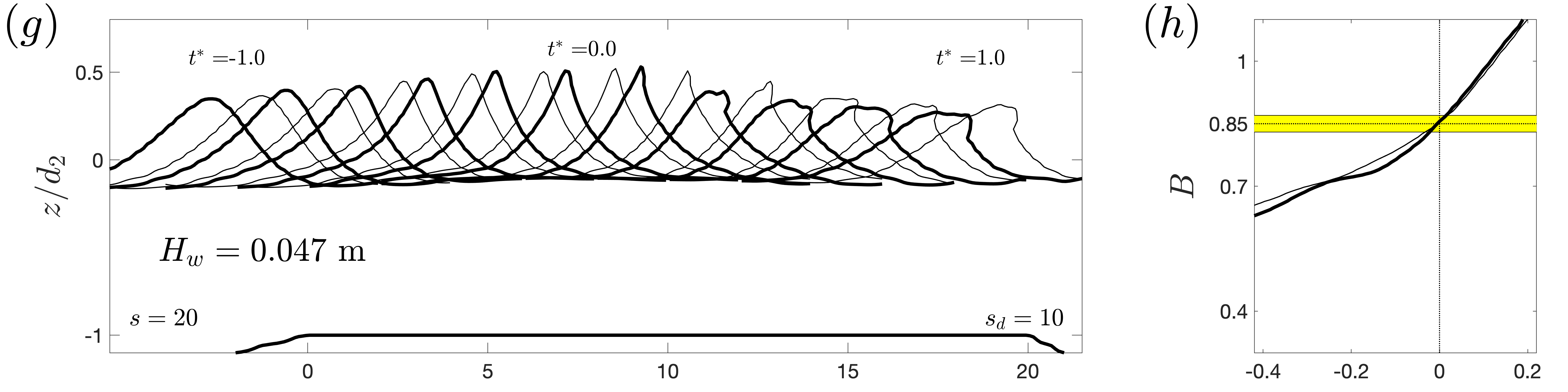}
\includegraphics[width=\textwidth]{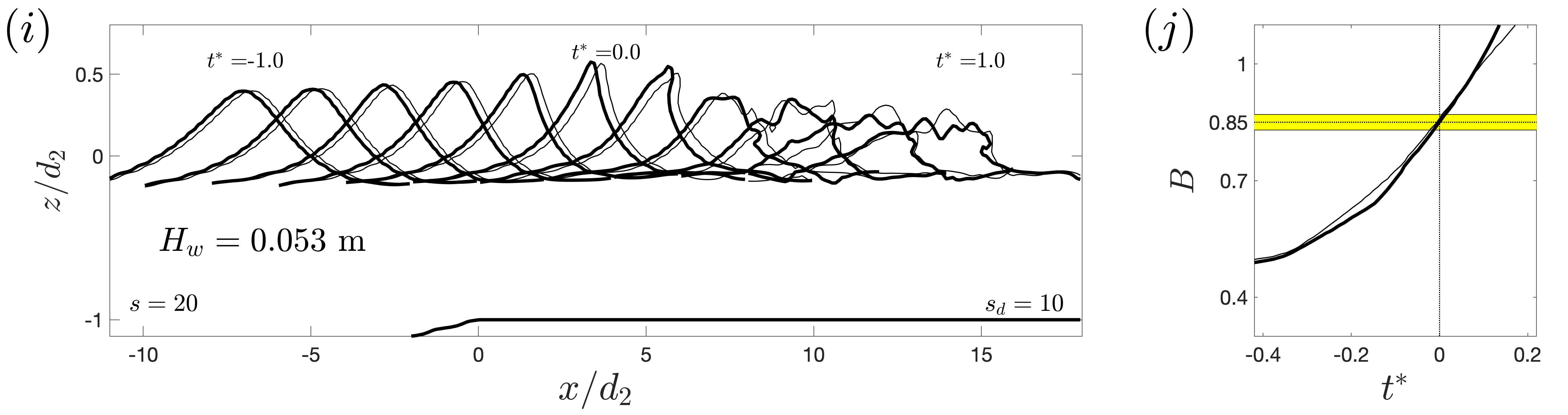}
\caption{Temporal evolution of $(a,c,e,g,i)$: wave profiles and $(b,d,f,h)$: the breaking onset parameter $B$ for two different evolving crests of a regular wave ($T_w = 1.01$ s) propagating over a bar with a front slope $m = 1/s$, demonstrating a transition from non-breaking to spilling breaking with an increasing $\xi_0$. Here $d_2$ is the still water depth over the top of the bar (Figure~\ref{fig1}$b$). All results are obtained using the LES/VOF model.}
\label{fig9}
\end{figure}

Figure~\ref{fig9} shows the temporal evolution of two evolving crests and their corresponding $B$ values for non-breaking and breaking regular waves ($T_w = 1.01$ s) propagating over a submerged bar, as defined in Figure~\ref{fig1}b.  Each row shows LES/VOF results for a case with an initial wave height $H_w$, where increasing $H_w$ results in a transition from non-breaking (Frames a-d) to intermittent breaking (Frames e and f) and breaking (Frames g-j) events. For each individual evolving crest, the reference time is the time at which $B$ passes $0.85$ or reaches its maximum for breaking and non-breaking cases, respectively. Although incident crests with the same $H_w$ have exactly the same initial wave conditions, their kinematics and dynamics near the break point or crest maximum are not the same, due to their interaction with the low-frequency waves in the numerical tank (e.g., seiches), the residual motions due to preceding waves, etc. Although these variations have a relatively small effect on the height of the evolving crests, they may result in an intermittent breaking, as shown in Frames (e) and (f). 

\begin{figure}
\centering
\includegraphics[width=\textwidth]{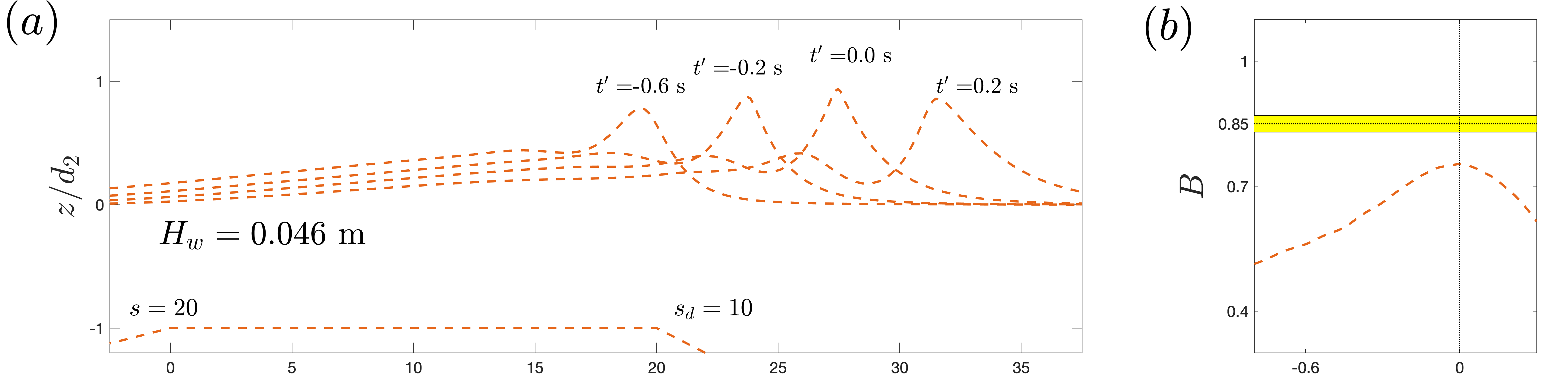}
\includegraphics[width=\textwidth]{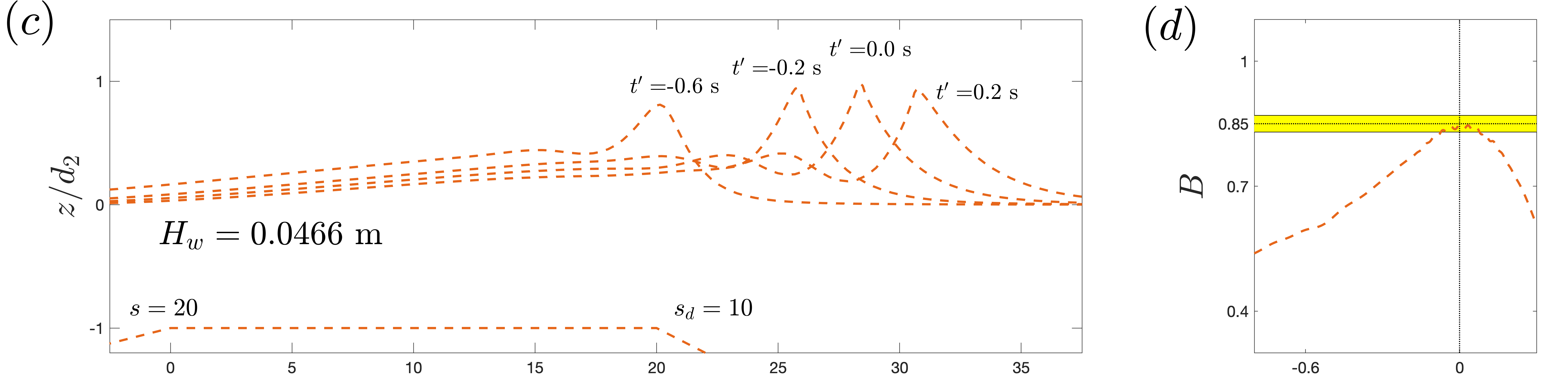}
\includegraphics[width=\textwidth]{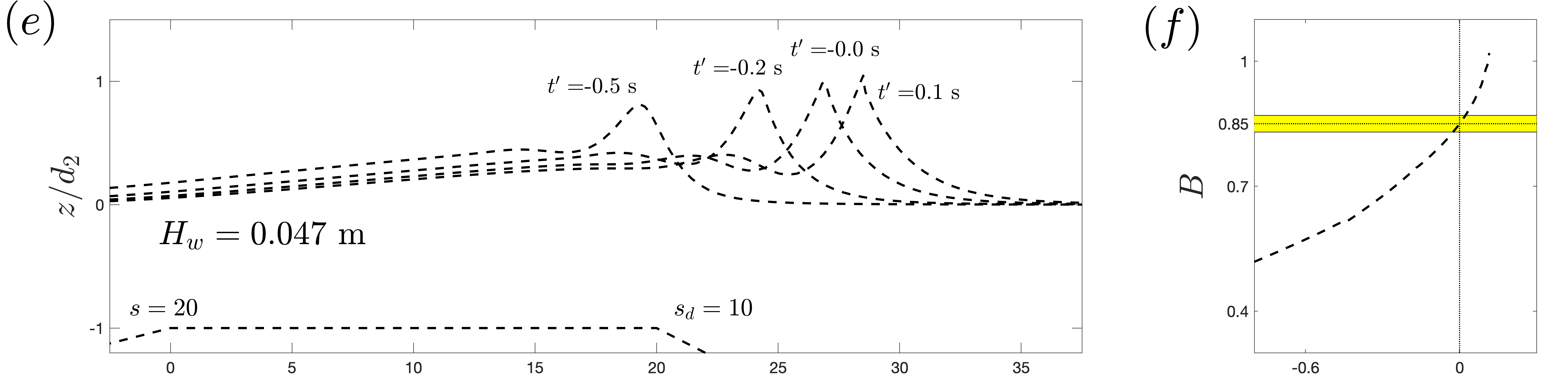}
\includegraphics[width=\textwidth]{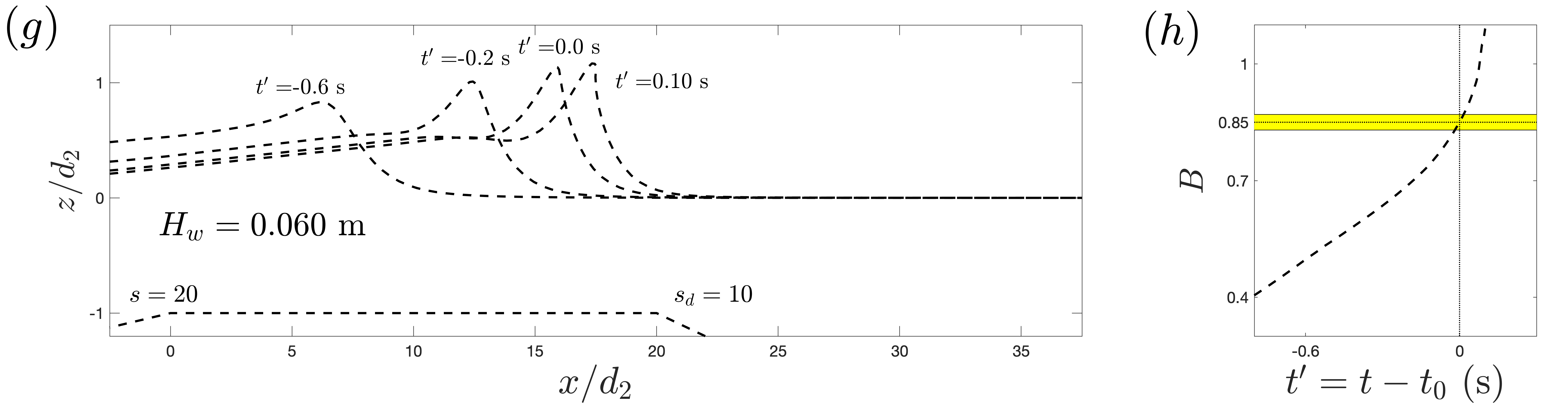}
\caption{$(a,c,e,g)$: Snapshots of the free surface elevations and $(b,d,f,h)$: temporal evolution of the breaking onset parameter $B$, for solitary waves propagating over a bar with a front slope $m = 1/s$ demonstrating a transition from non-breaking to spilling breaking with an increasing initial wave height. Here $d_2$ is the still water depth over the top of the bar (Figure~\ref{fig1}$b$).  All results are obtained using the BEM model.}
\label{fig10}
\end{figure}

Figure~\ref{fig10} shows similar results as in Figure~\ref{fig9} but for the solitary wave cases, computed using the BEM model.  Results shown in Figure~\ref{fig9} and Figure~\ref{fig10} confirm the validity of $B_{th}\approx0.85$ as a robust precursor of breaking onset in shallow water.  

\begin{figure}
\centering
\includegraphics[width=\textwidth]{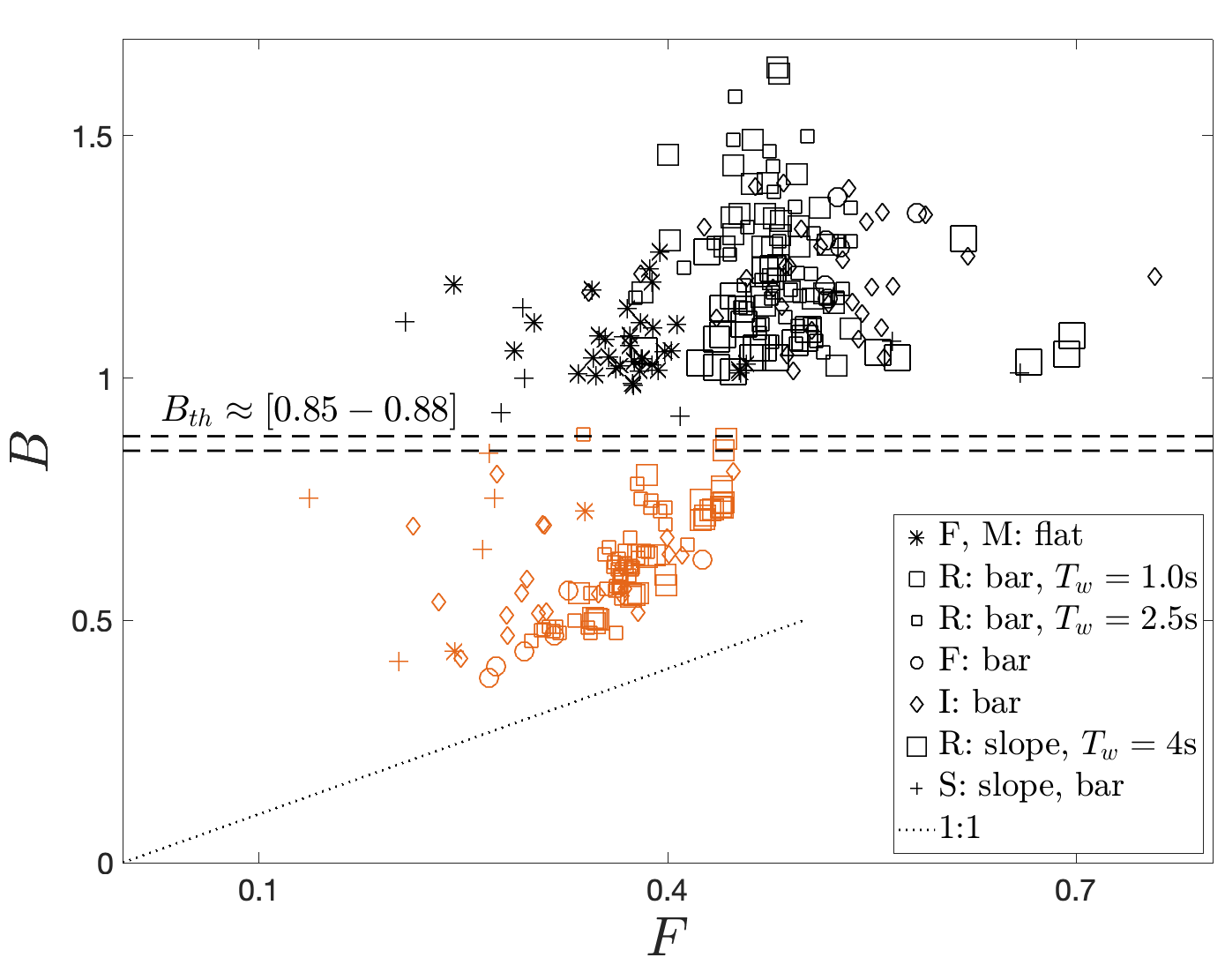}
\caption{Maximum value of the breaking onset parameter $B$ as a function of the wave Froude number $F$, for all breaking (black symbols) and non-breaking (orange symbols) wave crests. In the breaking cases, the maximum value of $B$ corresponds to the time, after the onset of breaking, at which the location of the crest maximum becomes noisy.}
\label{fig11}
\end{figure}

Finally, Figure~\ref{fig11} shows the variation of the maximum $B$ values as a function of the wave Froude number $F$ for all simulated crests, using LES/VOF and BEM models, from deep to shallow water. As mentioned above, we observe that if $B$ passes the threshold value $B_{th}\approx0.85$ it will also pass $1$ at the breaking onset for all cases. The two exceptional breaking wave cases indicated by + signs below $B=1$ represent solitary wave cases simulated using the BEM model, where the simulations stop before breaking onset due to insufficient spatial resolution.    
We observe that the threshold precursor values $B_{th}$, beyond which the crest evolves to breaking, range between 0.85 and 0.88 in shallow water wave breaking. This is consistent with the relevant previous studies of the variation of the parameter $B$ in intermediate depth and deep water \citep{Barthelemy-etal:2018,Saket-etal:2017,Saket-etal:2018,Derakhti-etal:2018}.  The plot also displays a dotted line corresponding to the linearized relation $B=F$.  We observe that the maximum occurring values of $B$ for all the tabulated steep but nonbreaking crests greatly exceed this lower limit, due to a combination of underprediction of fluid velocity in the crest as well as possible reductions of crest speed prior to breaking in intermediate depth cases.

\section{Discussion}
In this section, 
we present an evaluation of the other existing breaking criteria from the literature.  These are the various geometric parameters defined in \S\ref{sec3.3}, which are applied to the simulated wave trains.

\begin{figure}
\centering
\includegraphics[width=1\textwidth]{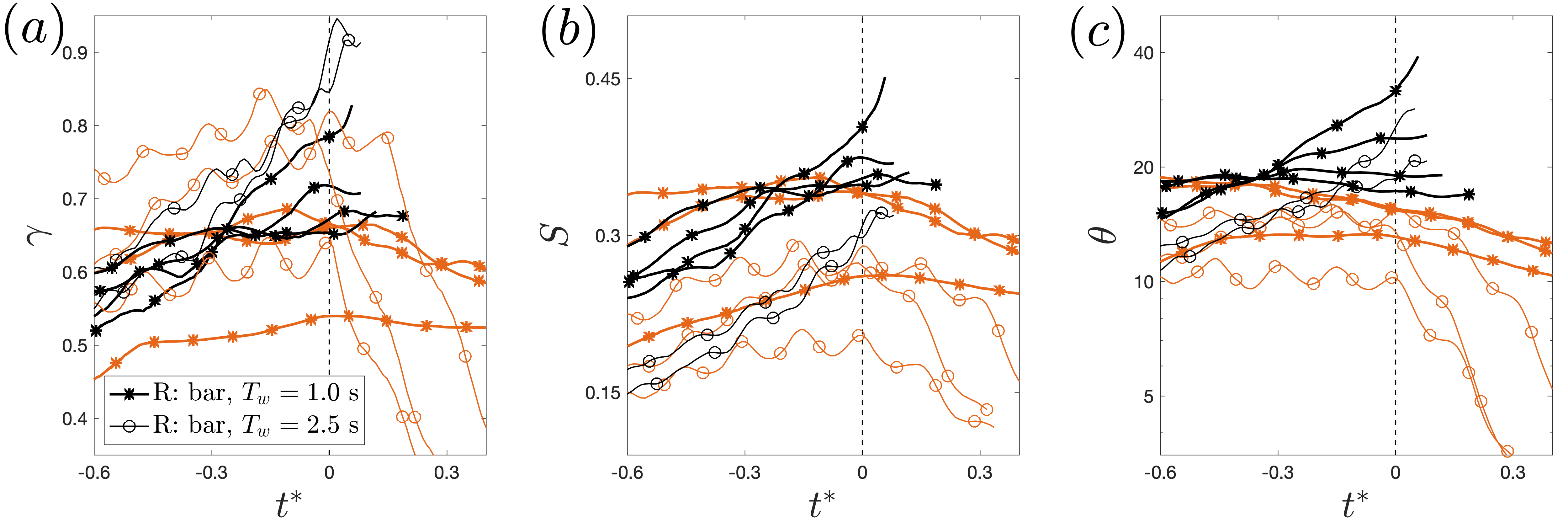}
\includegraphics[width=1\textwidth]{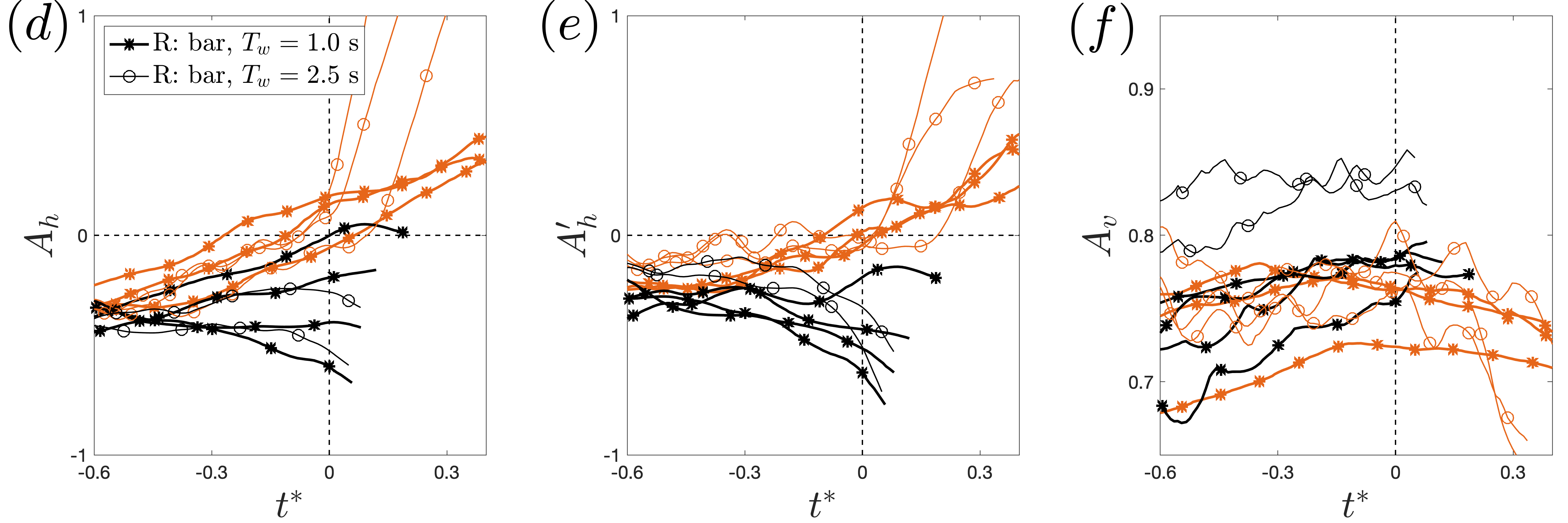}
\includegraphics[width=1\textwidth]{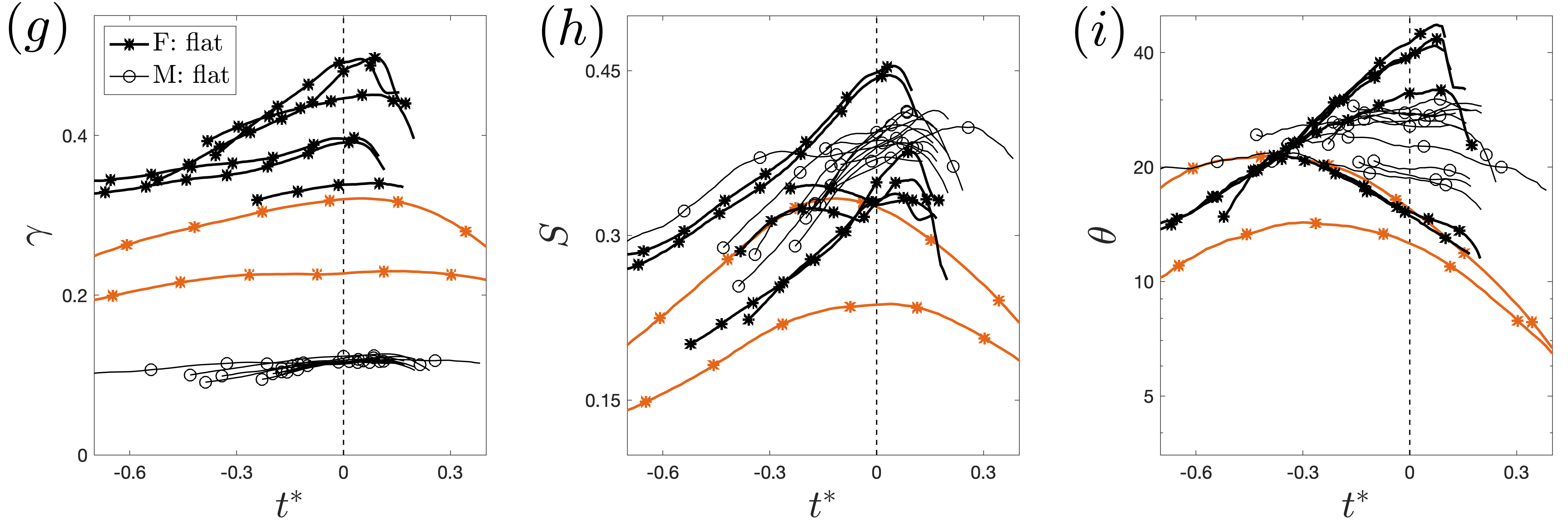}
\includegraphics[width=1\textwidth]{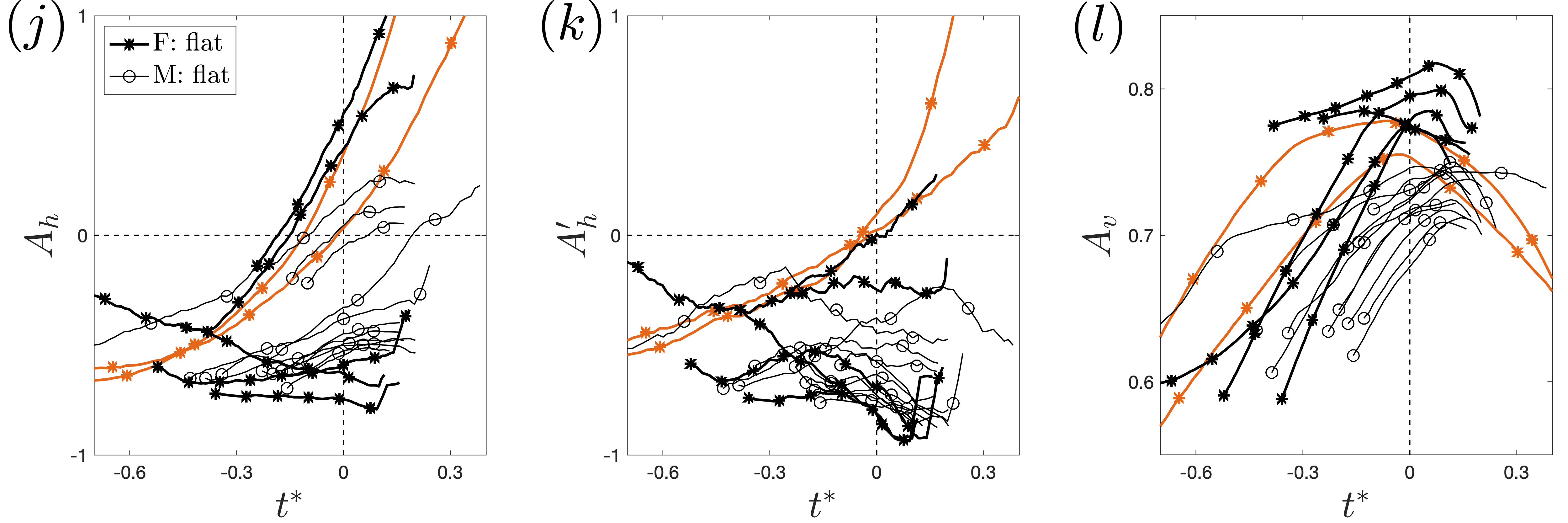}
\caption{Examples of temporal evolution of various geometric parameters defined in \S\ref{sec3.3} (also see Figure~\ref{fig2}) for breaking (black lines and symbols) and non-breaking (orange lines and symbols) wave crests $(a-f)$: in shallow water, and $(g-l)$: in intermediate depth and deep water. The capital letters in the legend indicate the type of incident waves, R: regular waves, F: focused packets, and M: modulated wave trains. In the legend, bar and flat denote bar geometry (Figure~\ref{fig1}$b$) and flat bed (Figure~\ref{fig1}$c$) respectively, and $T_w$ is the period of the regular incident waves. }
\label{fig3}
\end{figure}

\begin{figure}
\centering
\includegraphics[width=1\textwidth]{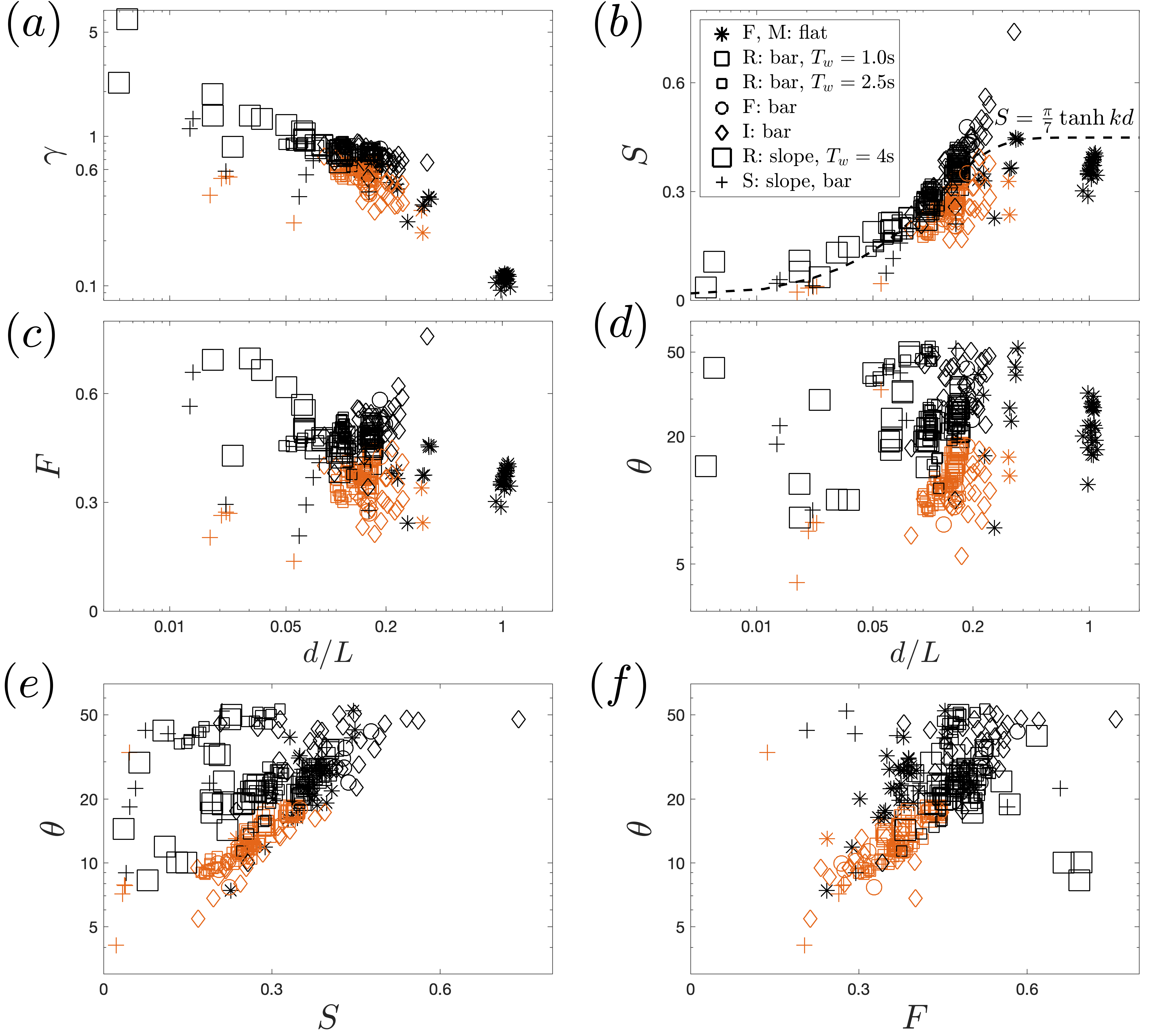}
\caption{Variation of various geometric parameters, defined in \S\ref{sec3.3}, at breaking onset or crest maximum, for all simulated breaking (black symbols) and non-breaking (orange symbols) wave crests from deep to shallow water. The capital letters in the legend refer to the type of incident waves, R: regular, I: irregular, S: solitary waves, F: focused packets, and M: modulated wave trains. Here, $\gamma = H/d$ is the nonlinear parameter (or breaking index), $S=\pi H/L$ is the wave steepness, $\theta= 180/\pi \tan^{-1}{(S^f)}$ is the wave front slope (all are defined in Figure~\ref{fig2}), and $F = ga/c^2_{lin}$ is the wave Froude number.}
\label{fig4}
\end{figure}

Figure~\ref{fig3} shows examples of computed temporal variation of the various geometric parameters defined in \S\ref{sec3.3} (also in Figure~\ref{fig2}) for breaking (black lines and symbols) and non-breaking (orange lines and symbols) wave crests from shallow to deep water. Examples shown in frames (a-f) represent regular waves shoaling over a submerged bar with a front face slope of $1/20$ (Figure~\ref{fig1}$b$), in which breaking is typically observed over the flat region of the bar and is characterized as shallow breaking. Examples shown in frames (g-l) represent focused packets and modulated waves propagating in intermediate and deep water over constant depth.  Further, Figure~\ref{fig4} shows variation of the four geometric parameters $\gamma,S,F$ and $\theta$ (\S~\ref{sec3.3}) at breaking onset or crest maximum, for which $t^* = 0$, for all simulated breaking (black symbols) and non-breaking (orange symbols) wave crests from shallow to deep water (which includes cases shown in Figure~\ref{fig3}).

The most commonly used breaking onset parameter in shallow water wave breaking is $\gamma = H/d$; in phase-averaged models the mean depth $d+\overline{\eta}$ is typically used instead of still water depth $d$, such that mean wave set-up or set-down is included. There is a large body of literature including laboratory and field studies attempting to define $\gamma$ values at breaking onset for various incident waves in shallow water. An extensive review is given in \citet{Robertson-etal:2013}. Observed values of $\gamma$ at breaking onset, in a wave-by-wave sense, are typically greater than 0.6 in shallow water. Consistent with the existing relevant literature, results shown in Figures~\ref{fig3}$a$ and \ref{fig3}$g$ indicate that 
$\gamma$ increases as a wave approaches the breakpoint and that $\gamma$ at breaking onset is an increasing function of the surf-similarity parameter $\xi_0$ \citep{battjes-icce74}. However, no unified formulation of $\gamma$ predicting the onset of depth-limited wave breaking can be found (see Figure~\ref{fig4}$a$). Further, it is clear that $\gamma$ is an irrelevant parameter for estimating the breaking onset of steepness-limited wave breaking in deep water.

In the shallow breaking cases shown in Figure~\ref{fig3}$a$, the local depth $d$ decreases over the front face of the bar (the shoaling region), then becomes constant over the top of the bar, and then increases over the back face of the bar (Figure~\ref{fig1}$b$). The latter explains the noticeable decrease of $\gamma$ for $t^*>0$ for non-breaking crests. During the time a non-breaking crest propagates over the top of the bar (constant depth region) the variation of $\gamma$ is relatively small.

Figures~\ref{fig3}$b$ and \ref{fig3}$h$ indicate that as a crest approaches breaking, or its maximum height for non-breaking crests, the local steepness $S$ (Eq.~\ref{eq11}) increases both in shallow and deep water cases. We observe that the maximum steepness values of all the simulated non-breaking crests are smaller than that given by \citet{Miche:1944}'s breaking steepness criterion $S = \pi/7\tanh{kd}$ (dashed line in Figure~\ref{fig4}$b$). We also observe that a large number of simulated breaking crests occur with a steepness value smaller than the limiting criterion. We note that our definition of $L$ is different from the classical definition for wavelength; our $L$ is much smaller than the latter in some of the shallow breaking cases considered here (see Appendix B).   In summary, breaking is clearly related to steepness, but a unified formulation that is able to predict maximum values of $S$ at breaking onset from deep to shallow water remains unknown; the same conclusion holds for the wave Froude number $F$ (Eq.~\ref{eq12}) (Figure~\ref{fig4}$c$).

Figures~\ref{fig3}c and \ref{fig3}i as well as Figure~\ref{fig4}d document the variation of the wave front slope $\theta$ (Eq.~\ref{eq13}) as a function of time and at $t^*=0$, respectively, from shallow to deep water. In general, breaking crests have higher maximum values of $\theta$ compared to non-breaking crests. However, most of the spilling breakers, both in deep and shallow water, maintain their maximum $\theta$ values as they approach the breakpoint. Moreover, $\theta$ decreases slightly as a crest approaches  breaking in marginal breaking cases, both in deep and shallow water. These observations suggest that $\theta$ might be a useful diagnostic breaking onset parameter but should be combined with other parameters; such as $\gamma$ in shallow and with $S$ (shown in Figure~\ref{fig4}e) in deep water or, more generally, with the wave Froude number $F$ (shown in Figure~\ref{fig4}f) in order to potentially predict the breaking onset time and location in a phase resolved sense.

Finally, frames (d) - (f) and (j) - (l) of Figure~\ref{fig3} demonstrate that neither the horizontal ($\mathcal{A}_h$ and $\mathcal{A}^{\prime}_h$) nor the vertical asymmetry of an evolving crest (as defined in \S~\ref{sec3.3}) are a good candidate as a breaking onset parameter. Further, results show that some of the simulated wave crests, both in shallow and deep water, are remarkably symmetric just prior to breaking ($\mathcal{A}_h\approx 0$). This result is consistent with field observations made using stereo photography in deep water \citep{schwendeman-thomson-jpo17} and with field observations using LIDAR in shallow water \citep{carini-phd-uw18}.

In summary, our results reveal that a criterion using both $\theta$ and $F$ has relatively higher skill in predicting the onset of breaking from deep to shallow water, compared to the other geometric parameters considered here. However, such a criterion still cannot segregate all breaking crests from non-breaking ones.

\section{Conclusions}\label{sec6}
The model simulation results presented here extend the results of \citet{Barthelemy-etal:2018} and \citet{Derakhti-etal:2018} to cases of waves shoaling and breaking in shallow water.  The local energy flux parameter $B$ exceeding the threshold of $\approx0.85$ is confirmed to provide a robust precursor to breaking for cases where breaking results from a crest instability.  In particular, we have simulated cases where a weak modulation of periodic waves by tank seiching leads to occasional breaking events in a train of otherwise unbroken waves, which are marginally close to breaking. These breaking events are clearly indicated by the passage of $B$ through the $\approx0.85$ threshold.  Further, we have shown that $B_{th}\approx 0.85$ clearly separates breaking and non-breaking cases for shoaling/de-shoaling waves propagating over bars. We conclude that this investigation provides further support for the generic applicability of the new breaking framework proposed by \citet{Barthelemy-etal:2018}, which was developed with specific reference to the onset of instability and incipient overturning in the region localized around wave crests.

Our extension to shoaling waves introduces the additional phenomenon of surging breakers, with breakdown and generation of turbulence during the uprush of a surging wave on a beach. This may be related more directly to instabilities of the strongly curved flow closer to the toe of the surging wave front.  This process is very different in nature from the mechanism covered by the analysis of \citeauthor{Barthelemy-etal:2018} and occurs without a crest-based criterion being exceeded.  It thus represents a different route to breaking whose occurrence (or onset) would require an alternate criterion to be developed.

The new criterion is suitable for use in wave-resolving models that cannot intrinsically detect the onset of wave breaking. Some of these models, such as High Order Spectral (HOS) models, become unstable if they reach the breaking onset stage, {\it i.e.}, $B=1$.
Thus, warning of imminent breaking onset at $B_{th}\approx 0.85$ is critical in the context of the successful application of the new criterion in such wave-resolving energy-conserving models; because at $B=B_{th}$ the waveform is well defined, no vertical tangent is developed on the wave front face, and the free surface is single-valued.

\vspace{0.5cm}
\noindent{\bf Acknowledgements}:  This work was supported by NSF Physical Oceanography grants OCE-1756040 and OCE-1756355 to the Universities of Washington and Delaware.  Computational support was provided by UD Instructional Technologies. MLB also gratefully acknowledges the support of the Australian Research Council for his breaking waves research through ARC DP120101701. SG gratefully acknowledge support from grant N00014-16-1270 from the US Office of Naval Research.

\appendix

\section{Model validation for shallow water breaking} \label{app:A}

In this section, the validation of the LES/VOF model \citep{Derakhti-Kirby:2014} including detailed comparisons of free surface evolution and organized and turbulent velocity fields, is presented for a number of available laboratory data for breaking and non-breaking waves in shallow water.  The reader is referred to \citet{Derakhti-Kirby:2014, Derakhti-Kirby:2014b, Derakhti-Kirby:2016} for the detailed examination of the model prediction of the free surface evolution, organized and turbulent velocity fields, bubble void fraction, integral properties of the bubble plume, and the total energy dissipation compared with corresponding measured data, as well as the sensitivity of the simulation results with respect to the selected grid resolution for focusing laboratory-scale breaking packets in intermediate depth and deep water.

In all the simulated cases using the LES/VOF model, the selected horizontal grid size in the wave propagation direction (which is always $+x$ direction here) $\Delta x$ is smaller than $1/100$ of the dominant wavelength at the $x$ location at which the crest maximum was observed, and $\Delta z  = \Delta y \le \Delta x$. Using such spatial resolution, our LES/VOF model captures the free surface and organized velocity field fairly accurately up to the break point, and the estimates of the loss of total wave energy due to wave breaking are typically within $10\%$ of observed levels \citep{Derakhti-etal:2018}, after correcting for the change in the downstream group velocity following breaking in isolated breaking waves \citep{Derakhti-Kirby:2016}. 

Regarding the FNPF-BEM model used in this work, \citet{grilli1994shoaling} showed that surface elevations simulated with the model for solitary waves shoaling over plane slopes agreed within $1-2\%$ with measured surface elevations, up to the breaking point.
\citet{grilli1994characteristics} reported a similarly good agreement of numerical results with experiments for solitary waves propagating over a trapezoidal breakwater. \citet{grilli1997breaking} showed that the model could accurately predict  breaking crest elevations, breaker index, and breaker types  for solitary waves of various incident height propagating over mild to steep slopes. Finally, \citet{grilli2019fully} show that the model also accurately simulates the shoaling and propagation of periodic waves over a bar similar to that considered here.

\subsection{Regular waves shoaling over a plane beach}
Here we consider the LES/VOF model performance for the case of regular depth-limited wave breaking on a planar beach (P10-r) in terms of phase-averaged free surface elevations and wave height using the data set of \citet{ting-nelson-ceng11}. 
We also compare the model results of the case P10-r with the free surface and velocity measurements of the spilling case of \citet{Ting-Kirby:1994}. The experimental set-up and incident wave conditions of the latter are similar as in P10-r and are also summarized in Table \ref{tab:A1}. This experiment has been widely used by other researchers to validate both RANS \citep{lin-liu-ww99, ma-etal-jgrc11, Derakhti-etal:2015,Derakhti-etal:2016a, Derakhti-etal:2016b, Derakhti-etal:2016c} and LES \citep{Christensen:2006, Lakehal-Liovic:2011} numerical models.

\begin{figure}
\centering
\includegraphics[width=\textwidth]{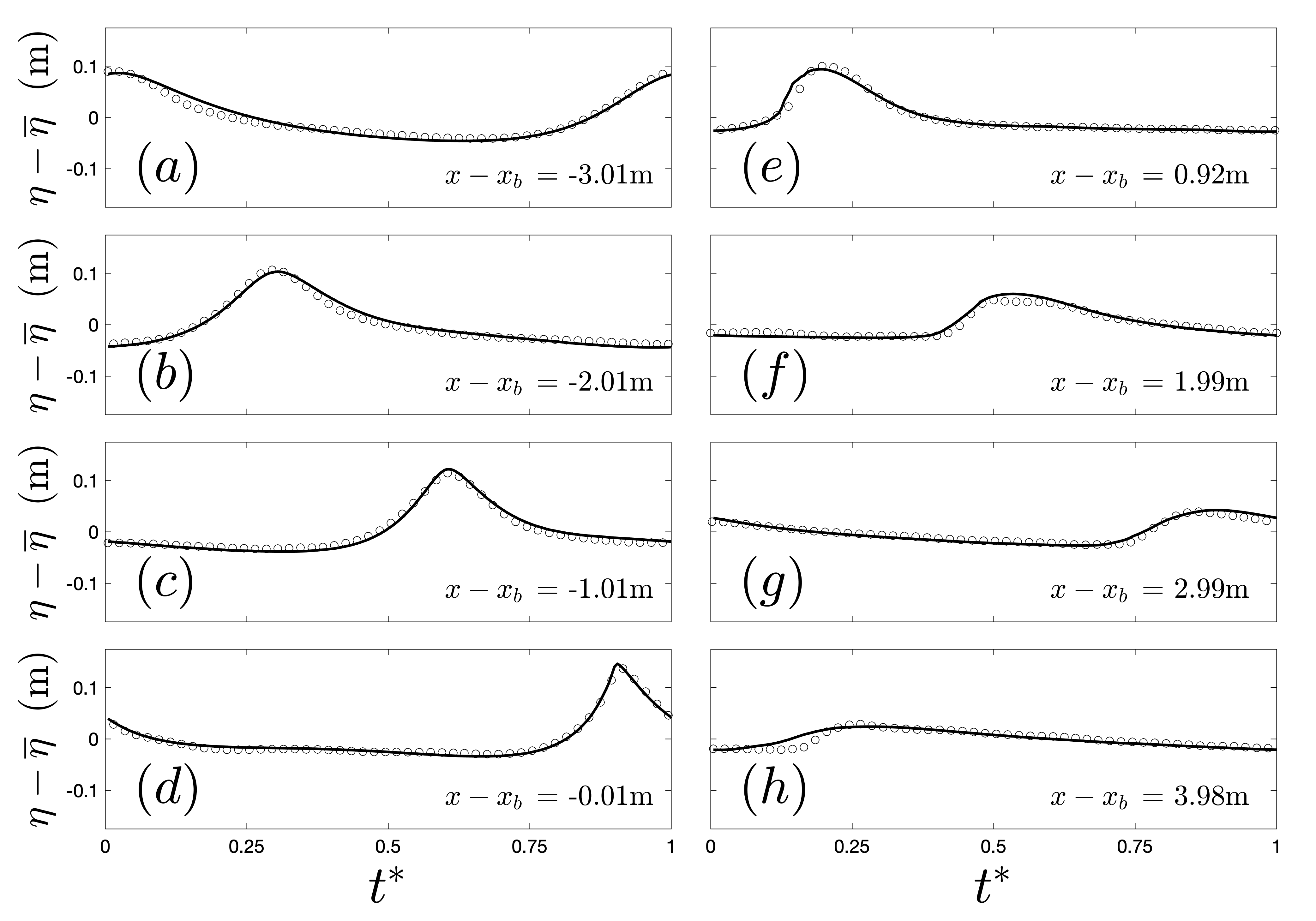}
\caption{Comparison between the LES/VOF model results of spanwise-phase-averaged free surface elevations at various cross-shore locations for the case P10-r and the corresponding measurements by \citet{ting-nelson-ceng11}. 
No spanwise averaging was involved in the measurement.}
\label{fig3.1}
\end{figure}

\begin{figure}
\centering
\includegraphics[width=\textwidth]{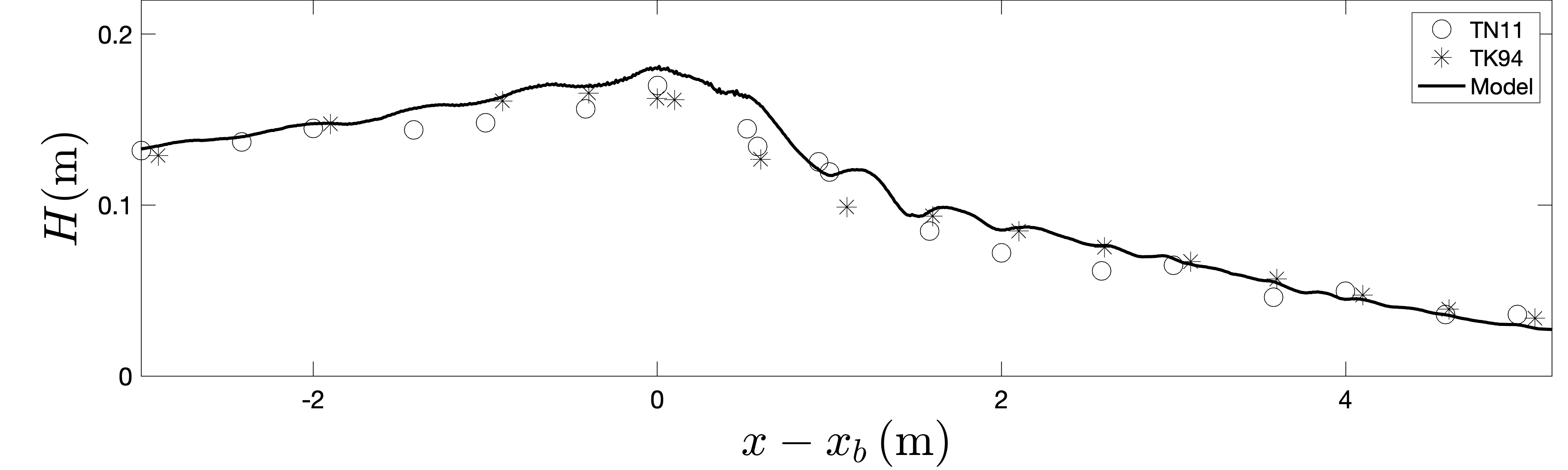}
\caption{The LES/VOF model-data comparison of the cross-shore variation of the wave height $H$ for the case P10-r. Here TN11 and TK94 denote the data set of \citet{ting-nelson-ceng11} and \citet{Ting-Kirby:1994} respectively.}
\label{fig3.2}
\end{figure}

Figure \ref{fig3.1} shows that the model captures the evolution of phase-averaged free surface elevations reasonably well compared with the corresponding measurements of \citet{ting-nelson-ceng11} in the shoaling, transition and inner surfzone. 
Further, Figure \ref{fig3.2} shows the comparison between the predicted and observed cross-shore variation of the wave height $H$ calculated from the phase-averaged free surface time-series. Here phase averaging is performed over $N$ successive waves after the wave field reaches a steady state condition, where $N$ is 10 in both the simulated results and the measurements.

\begin{table}[h]
\centering
\begin{tabular}{c c c c c c c c c c c c c}
     Case &$H_w$				&$T_w$ 	&$d_1$ &$L_1$ &$s$ &$\xi_0$ &$d_2$ &$L_2$  & $s_d$ & Exp. \\[3pt]
     	      &	(mm)			&(s) 	&(m)   &(m)   &    &       &(m)   &(m)       &  \\ [3pt]
\hline
     P10-r	  &122				&2.0	&0.36  &0     &$\frac{100}{3}$  &0.21   &-   &-         & - & \citet{ting-nelson-ceng11}\\[3pt]     
        	  &125				&2.0	&0.4  &0     &$35$  &0.20   &-   &-         & - & \citet{Ting-Kirby:1994}\\[3pt]     
     B1-r	  &41.0 &1.01	&0.4   &6     &20  &0.30        &0.1   &2        &10 & \citet{Luth-etal:1994:kinematics}\\[3pt]     
     B3-r	  &29.0 &2.53	&0.4   &6     &20  &0.95        &0.1   &2       & 10 & \citet{Luth-etal:1994:kinematics}\\[3pt]     
     B9-r	  &97.2				&1.43	&0.7   &2     &10  &  0.57      &0.08   &0      & 0  & \citet{Blenkinsopp-Chaplin:2007}\\ [3pt] 
\hline
\end{tabular}
\caption{Input parameters for the simulated cases used for the validation of the LES/VOF model. Definitions are given in table \ref{table1}.}
\label{tab:A1}
\end{table}

Figure \ref{fig3.2} also shows that the spatial evolution of $H$ relative to the break point in the case P10-r is comparable with that in the spilling case of \citet{Ting-Kirby:1994}. Thus although the incident wave conditions and setup in the latter are slightly different than those in the case P10 the wave-driven currents and turbulence statistics should be comparable.

\begin{figure}
\centering
\includegraphics[width=0.49\textwidth]{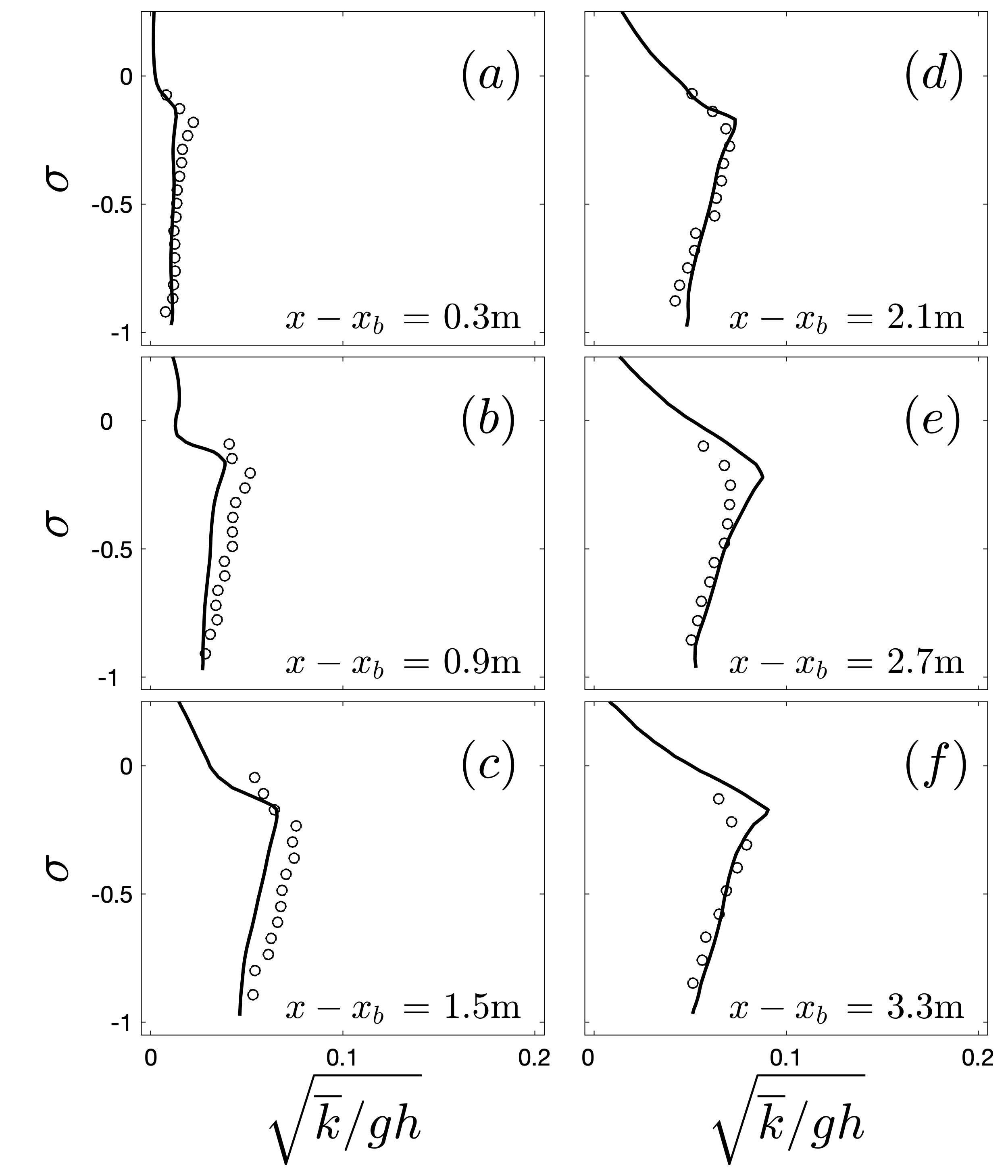}
\includegraphics[width=0.49\textwidth]{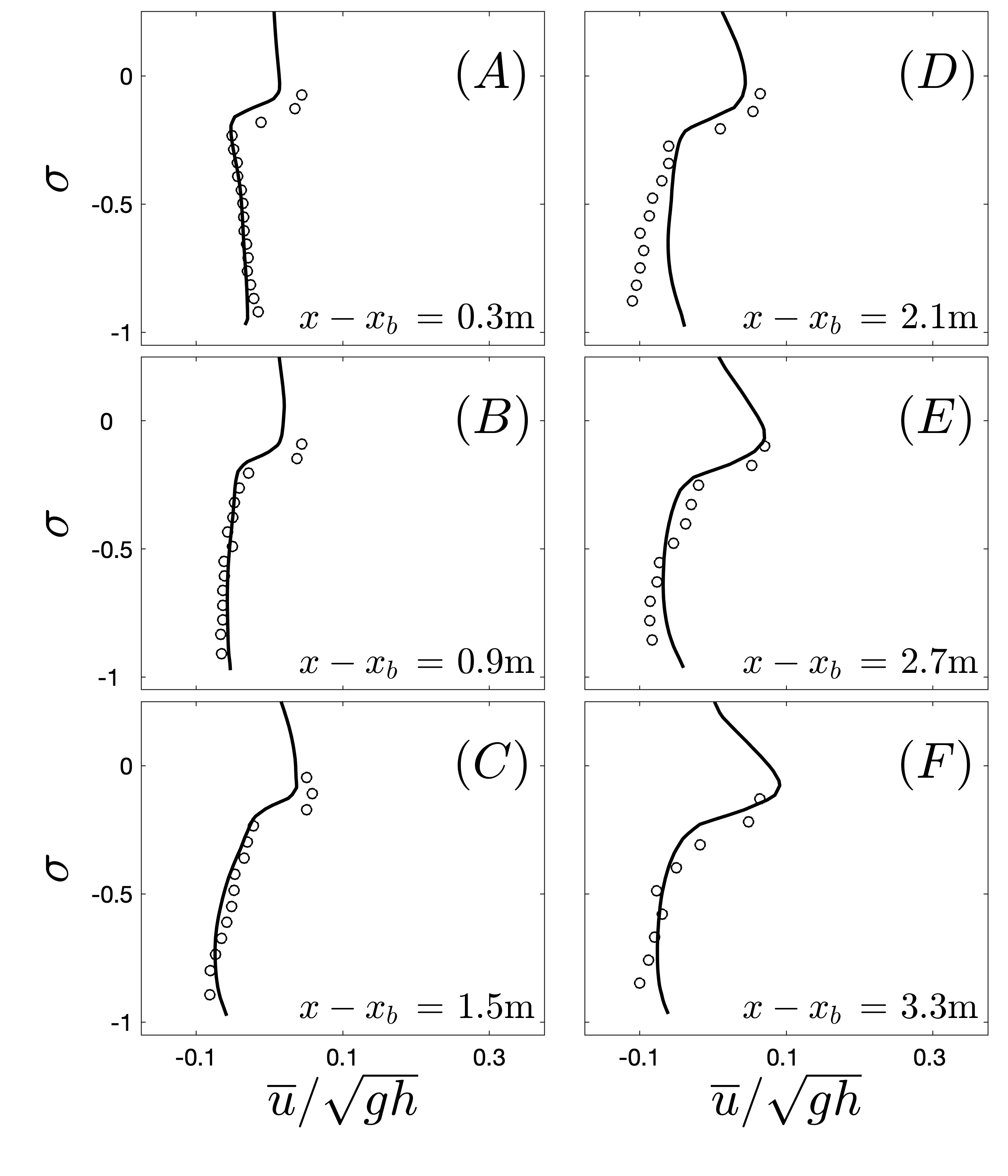}
\caption{The LES/VOF model results of spanwise-time-averaged normalized $(a-f)$ turbulent kinetic energy, $\sqrt{\overline{ k}/{gh}}$, and $(A-F)$ horizontal velocity, $\overline{ u }/\sqrt{gh}$, (undertow) profiles for the case P10-r at various cross-shore locations after the initial break point. Circles show the measurements of \citet{Ting-Kirby:1994}. Here, $\sigma=(z-\overline{\eta})/h$ and $h = d+\overline{\eta}$.}
\label{fig3.3}
\end{figure}

Figure \ref{fig3.3} shows the spatial distribution of the normalized spanwise-time-averaged, $\overline{\langle k \rangle}^{1/2}/\sqrt{gh}$, turbulent kinetic energy for PS-a. 
Figure \ref{fig3.3} shows that both the magnitude and spatial variation of the predicted $\overline{\langle k \rangle}^{1/2}/\sqrt{gh}$ and $\overline{\langle u \rangle}/\sqrt{gh}$ are consistent with the corresponding measured values of \citet{Ting-Kirby:1994} in the transition and inner surf zone.

\subsection{Regular waves shoaling over an idealized bar}

\begin{figure}
\centering
\includegraphics[width=1.0\textwidth]{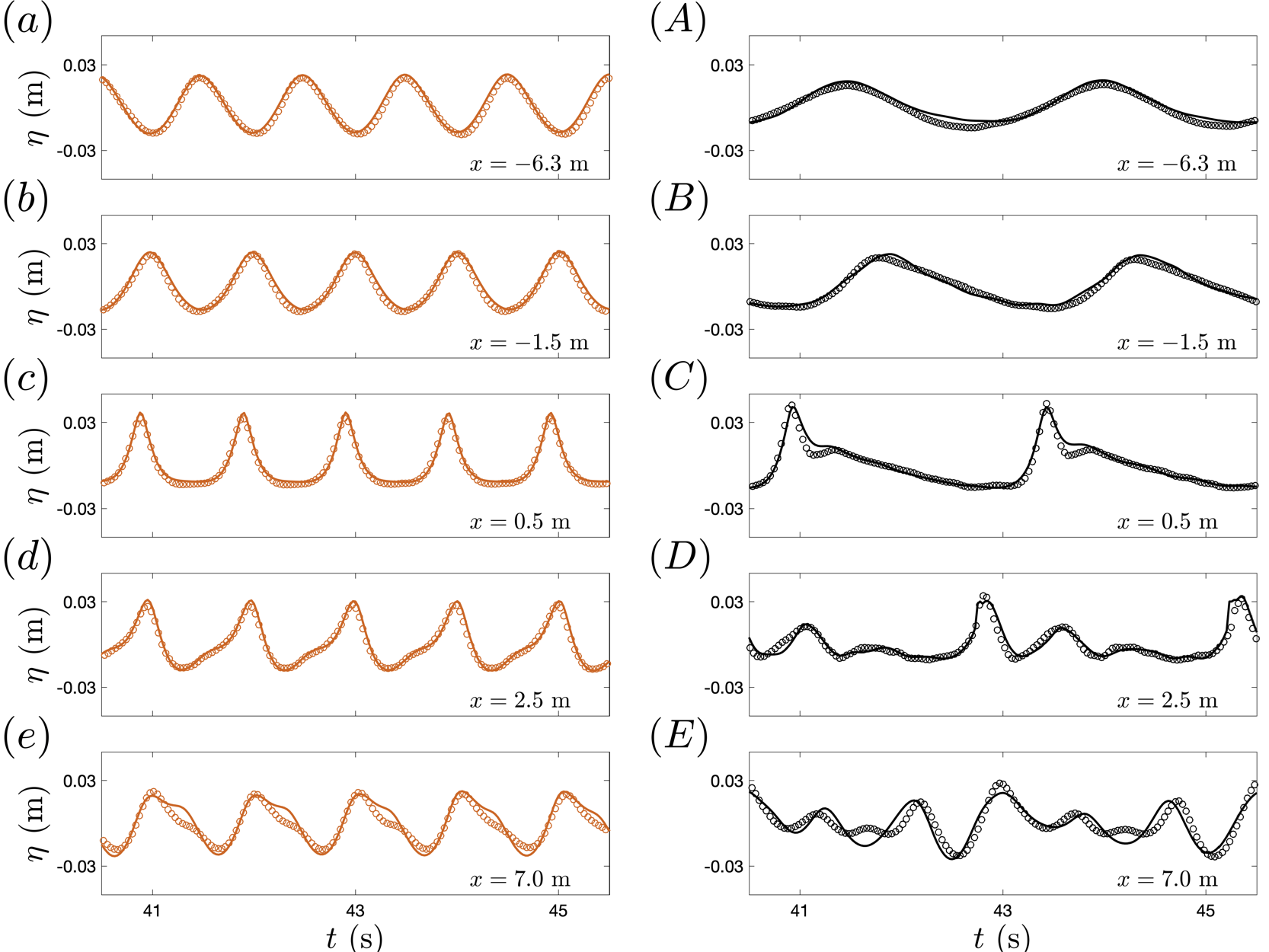}
\caption{Comparison of the LES/VOF model results (solid lines) and measurements \citep{Luth-etal:1994:kinematics} (circles) of free surface elevations at various $x$ locations for the along-crest uniform $(a-e)$ non-breaking, with $T_w = 1.01$ s and $H_w = 0.041$ m, and $(A-E)$ breaking, with $T_w = 2.525$ s and $H_w = 0.029$ m, regular waves shoaling over a submerged bar. Here $-6<x<0$ and $0<x<2$ indicate the up-slope and top of the bar respectively (see Figure~\ref{fig1}c and Table~\ref{tab:A1}).}
\label{fig3.4}
\end{figure}

\begin{figure}
\centering
\includegraphics[width=1.0\textwidth]{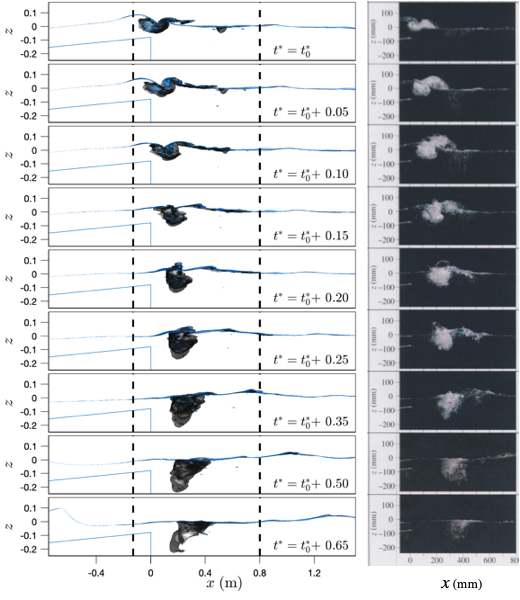}
\caption{Comparison of the side-view of the predicted (left column) and observed (right column) bubble plume evolution for the case B9-r. The two dashed lines in the right column indicate the field of view of the photographs, adopted from \citet[][Figure 4]{Blenkinsopp-Chaplin:2007}.}
\label{fig3.5}
\end{figure}

Here we consider the LES/VOF model performance for cases of regular non-breaking (B1-r) and breaking (B3-r and B9-r) waves shoaling over a submerged bar, using the data sets of \citet{Luth-etal:1994:kinematics} and \citet{Blenkinsopp-Chaplin:2007}. Figures~\ref{fig3.4} and \ref{fig3.5} documents that the model accurately captures the nonlinear evolution of evolving crests propagating over the up-slope ($-s(d_1-d_2)<x<0$) and top ($0<x<L_2$) of the bar in all cases. Figure~\ref{fig3.5} also shows that the model fairly reasonably predicts the kinematics of the entrained bubble plume compared to the observations. The apparent mismatch between the predicted and observed wave profiles is mainly due to the mismatch between their corresponding incident waves and due to the difference between the low frequency wave climate in the numerical and laboratory wave tanks. 

\section{Sensitivity of the local geometric parameters of an evolving crest} \label{app:B}
Two main sources of uncertainty in the value of the geometric parameters defined in \S~\ref{sec3.3} are the selected definitions of the local length $L$ and height $H$ of an evolving crest. Here we quantify such uncertainties in detail.

As mentioned in \S~\ref{sec3.3}, we define the local wave length $L =$ Min$(2l_{zc},2l_{zc}^{sg})$, where $l_{zc}$ (see Figure~\ref{fig2}) is the distance between the two consecutive zero-crossing points adjacent to the crest. Here $l_{zc}^{sg}$ is a length scale obtained from the skewed-Gaussian fit $f(r)$ defined as a scaled product of the standard normal probability density function $\phi(r) = \exp{[-r^2/2]}/\sqrt{2\pi}$ and its cumulative distribution function $\Phi(r) = (1+\text{erf}{[r/\sqrt{2}]})/2$ (\text{erf} denotes the error function) given by
\begin{equation}
    f(r) = c_1\phi(r)\Phi(\alpha r) + c_2, \label{SG}
\end{equation}
where $r = (x-x_p)/\omega$ with $x_p$ and $\omega$ are the peak location and scale respectively, $\alpha$ the horizontal skewness parameter ($\alpha<0$ for waves pitch forward), $c_1$ a scaling parameter and $c_2$ a vertical offset. The instantaneous $f$ for each crest is obtained by a nonlinear fitting of Eq.~\ref{SG}, including five coefficients, to the corresponding simulated wave profile. 

\begin{figure}
\centering
\includegraphics[width=1\textwidth]{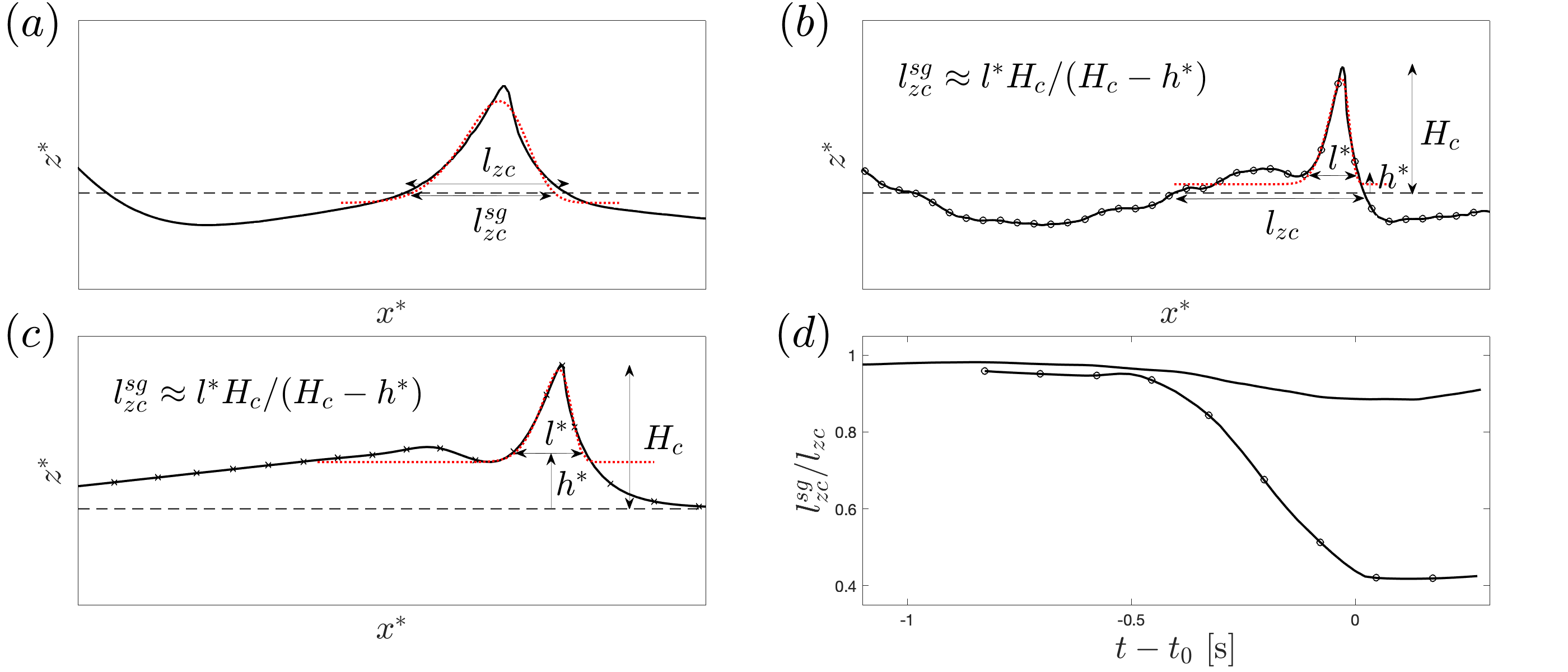}
\caption{$(a,b,c)$ Definition of the local zero-crossing length-scale $l_{zc}^{sg}$ obtained from skewed-Gaussian fitting (dotted lines) to the wave profile (solid lines) for examples of evolving crests shoaling over a submerged bar as well as $(d)$ the temporal variation of $l_{zc}^{sg}/l_{zc}$ before (shoaling phase) and after the breaking onset ($t = t_0$) for the crests shown in $(a)$ and $(b)$. $(a)$ Regular waves with $T_w = 1.01$ s, $(b)$ regular waves with $T_w = 2.525$ s, and $(c)$ a solitary wave. Note that $l_{zc}$ does not exist for solitary waves. In $(a,b,c)$, the dashed lines show the still water levels. }
\label{figB1}
\end{figure}

Frames (a), (b), and (c) of Figure~\ref{figB1} show examples of $f$ (dotted lines) and the corresponding $l_{zc}^{sg}$, just before breaking onset time, for three simulated evolving crests shoaling over a submerged bar. In addition, Figure~\ref{figB1}$d$ shows the temporal variation of the ratio $l_{zc}^{sg}/l_{zc}$ for the two examples shown in frames (a) and (b). Frames (b) and (d) show that we may have $l_{zc}^{sg}\ll l_{zc}$ at breaking onset in cases with irregularities on the back face of the wave, {\it e.g.}, due to the presence of higher harmonics.
Finally, in solitary cases (Figure ~\ref{figB1}$c$) we simply define $L = 2l_{zc}^{sg}$ because there are no zero-crossing points and thus $l_{zc}$ cannot be defined.

\begin{figure}
\centering
\includegraphics[width=0.51\textwidth]{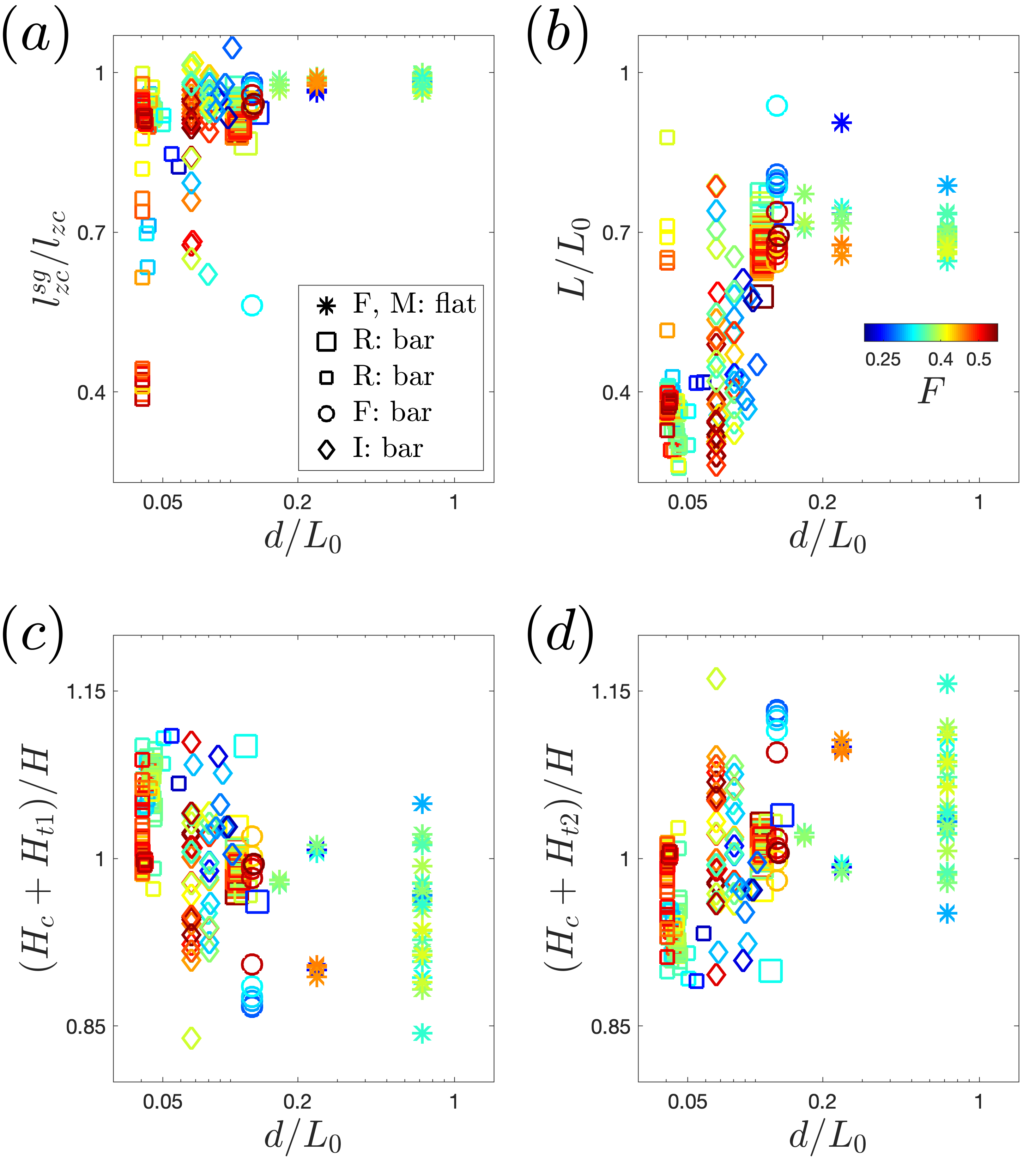}
\caption{Sensitivity of the local height and length of an evolving crest from deep to shallow water. $(a)$ the ratio between the length scales $l_{zc}^{sg}$ obtained from skewed-Gaussian fitting defined in Eq. (\ref{SG}) and $l_{zc}$ both shown in Figure~\ref{figB1}; and $(b)$ the ratio between the zero-crossing length scale $L =$ Min$(2l_{zc},2l_{zc}^{sg})$ and a wave length $L_0$ at breaking onset for the breaking crests or at the time at which $H_c = \eta_{max}$ for the non-breaking crests. Here, $L_0$ is obtained by using the linear dispersion relation $(2\pi/T_0)^2 = gk_0\tanh{[k_0d]}$ with $d$ the still water depth, $k_0 = 2\pi/L_0$ and $T_0$ equals to paddle period for monochromatic waves and peak period $T_p$ for incident irregular waves.}
\label{figB2}
\end{figure}

At breaking onset, Figure~\ref{figB2}$a$ demonstrates that the length scale $l_{zc}^{sg}$ obtained from the skewed Gaussian fitting (Eq. \ref{SG}) is usually smaller than the zero-crossing length scale $l_{zc}$ (Figure~\ref{figB1}). Our results show that $l_{zc}^{sg}/l_{zc}>0.9$ in most cases, especially for those with $d/L_0>0.1$, with $d$ the still water depth and $L_0$ a linear prediction of the local wave length obtained by using the linear dispersion relation $(2\pi/T_0)^2 = gk_0\tanh{[k_0d]}$ with $d$ the still water depth, $k_0 = 2\pi/L_0$ and $T_0$ equal to paddle period for monochromatic waves and peak period $T_p$ for incident irregular waves. In some of the shallow cases ($d/L_0<0.1$), however, we observe $l_{zc}^{sg}/l_{zc}$ values down to 0.4. 
Figure~\ref{figB2}$b$ shows that our definition of $L$ represents a smaller length scale compared to the characteristic wave length $L_0$ where the averaged values of $L$ vary between $L_0/3$ in shallow water up to $0.7L_0$ in intermediate and deep water. 

Further, our results (Figures~\ref{figB2}$c$ and ~\ref{figB2}$d$) indicate that other potential definitions of wave height such as $H_c+H_{t1}$ or $H_c+H_{t2}$ are within 10$\%$ of $H = H_c+(H_{t1}+H_{t1})/2$ in most cases from deep to shallow water. In addition, the downstream trough height $H_{t1}$ is greater than or equal to the upstream trough height $H_{t2}$ in shallow water cases; the trend is reversed in deep water cases. These trough heights vary between $0.2H_c$ and $0.5H_c$ in most cases.

\clearpage
\bibliographystyle{jfm.bst}
\bibliography{refs}

\end{document}